\documentclass[showpacs,amsmath,amssymb,twocolumn,aps]{revtex4} 
\usepackage{graphicx}
\usepackage{epsfig}
 \usepackage{pstricks}
\usepackage{psfrag}
\usepackage{dcolumn}
\usepackage{bm}

 \def\comment#1{}
\def\mn#1{*{\marginpar{\footnotesize #1}}}
\def\mn#1{}

\begin{document}
\title{Phase Diagram of Vortices
in High-${\bf  T_c} $ Superconductors with a Melting Line 
in the deep $ H_{c2} $ Region} 
\author{J\"urgen Dietel}
\affiliation{Institut f\"ur Theoretische Physik,
Freie Universit\"at Berlin, Arnimallee 14, D-14195 Berlin, Germany}
\author{Hagen Kleinert}
\affiliation{Institut f\"ur Theoretische Physik,
Freie Universit\"at Berlin, Arnimallee 14, D-14195 Berlin, Germany}
\affiliation{ICRANeT, Piazzale della Repubblica 1, 10 -65122, Pescara, Italy}

\date{Received \today}
\begin{abstract}
We use a simple 
elastic Hamiltonian for the  vortex lattice in a weak  impurity background
which includes defects  in the form of integer-valued fields
to calculate the free energy of a vortex lattice  
in the deep $ H_{c2} $ region. The phase diagram 
in this regime  is obtained by applying 
the variational approach of M{\'e}zard and Parisi
developed for random manifolds.
We find a first-order line between the Bragg-glass and 
vortex-glass phase as a continuation of the melting line.  
In the liquid phase, we obtain 
an almost vertical third-order glass transition line near the 
critical temperature in the $ H-T $   plane.  
Furthermore, we find an almost vertical 
 second-order phase transition line in the Bragg-glass as 
well as the vortex-glass phases which crosses the   
first-order Bragg-glass, vortex-glass transition line. We calculate the jump 
of the temperature derivate 
of the induction field across this second-order line as well as 
the entropy and magnetic field jumps across the first-order line. 
\end{abstract}

\maketitle

\section{Introduction}
 
The phase diagram of high-$ T_c $ superconductors as a function of the 
magnetic field $ H $ and temperature $ T $   
is mainly governed  by the interplay of thermal fluctuations and 
quenched disorder \cite{Blatter1,Nattermann1}, leading to  
various different states of the vortex matter summarized in Fig.~1.  

\begin{figure}[t]
 \begin{center}
   \includegraphics[height=6cm,width=8cm]{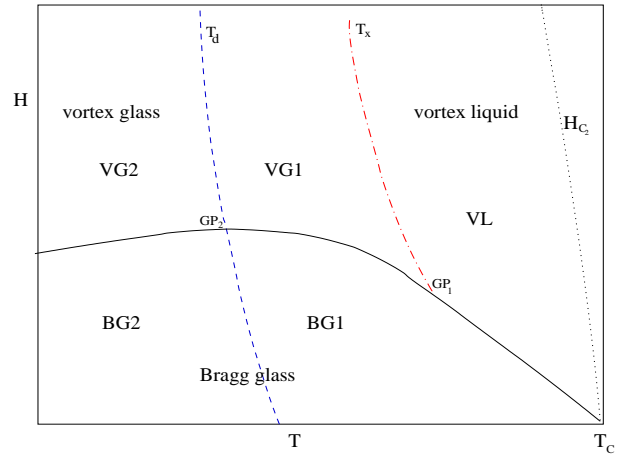}
 \end{center}
 \caption{Sketch of the phase diagram of BSCCO or similar
high-$T_c$ superconductors where the phase transition lines lie far below 
$ H_{c2} $. The solid line denotes a first-order phase transition
line being a first-order  melting transition between the BG1-VL, 
BG1-VG1 phase and a disorder induced first-order transition  
between the BG2-VG2 phase. The (blue) dashed curve denoted by $ T_d $ 
is found to be  a second-order glass transition line
\cite{Beidenkopf2}. The intersection point with the first-order line 
is denoted by GP2. The (red) dashed-dotted  curve 
is the $ T_x $ line found by Fuchs 
{\it et al.}. \cite{Fuchs1} using  surface barrier experiments. It intersects 
the first-order line in the point GP1. 
 }
 \end{figure}
At low magnetic field the vortex solid melts into a 
vortex liquid (VL) via a first-order melting transition.  
Prominent examples of high-$T_c$ superconductors exhibiting
a solid-liquid  melting transition are the
anisotropic compound
$ {\rm YBa}_2 {\rm Cu}_3 {\rm O}_{7-\delta} $ (YBCO),
 and the strongly layered
compound $ {\rm Bi}_2 {\rm Sr}_2 {\rm Ca}
{\rm Cu}_2 {\rm O}_8 $ (BSCCO).
The position of the melting line in the $ H-T $ plane 
is mainly influenced by  the anisotropy of the superconductor 
\cite{Dietel2}. In   YBCO with its low anisotropy, most of the melting line 
$ B_m$ lies in the vicinity of the upper critical field 
$ H_{c2} $, i.e. $ B/H_{c2} \gtrsim  0.5 $ where $ B \approx H $ 
is the induction 
field in the relevant regime. The phase diagram 
for superconductors with a melting line in this regime was 
discussed by us in Ref.~\cite{Dietel3}.  
This magnetic field regime consists of a 
vortex lattice, or vortex fluid, with overlapping vortex cores,
if we use elastic constants calculated in Refs. \onlinecite{Brandt1,Hetzel1}. 
The  phase diagram was derived on the basis of a defect 
melting model set up in Refs. \cite{GFCM2, Dietel1}. 
The model is Gaussian in the 
elastic strains and takes into account the defect degrees of freedom 
by integer-valued gauge fields. From this we derive effective 
Hamiltonians for the low-temperature solid and high-temperature liquid  
phases by summing over all defect fields. By further integrating 
out vortex degrees of freedom we obtain the partition functions of both 
phases. This is done with the help of the variational approach 
of M{\'e}zard and 
Parisi \cite{Mezard1}, originally developed for random manifolds and applied 
later to vortex lattices without defects in Refs. 
\onlinecite{Giamarchi1, Korshunov1}. A similar approach was used 
recently to calculate the glass transition line for YBCO 
via Ginzburg-Landau theory \cite{Li1}.  

When including weak pinning, the solid phase becomes
a  quasi-long-range
ordered Bragg-glass  \cite{Blatter1}.
At higher
 magnetic fields, the quasi-long-range order is destroyed
and there exist also a vortex-glass phase.
The transition is marked by
 the disappearance of Bragg peaks in scattering data.
We obtain in Ref.~\onlinecite{Dietel3} a phase diagram 
consisting of a unified first-order phase 
line between the Bragg-glass phase and 
the vortex-glass phase and the vortex-liquid which is 
sketched for BSCCO in Fig.~1. We point out that the first-order 
character of the transition line between the Bragg-glass phase 
 and the vortex-glass phase is not 
experimentally established yet for YBCO. It is deduced from  
magnetic anomalies in response to the external magnetic field.  
For BSCCO, the first-order character 
of the Bragg-glass, vortex-glass line was measured  by 
supercooling  \cite{Beek1} 
and magnetic field shaking techniques \cite{Avraham1}. 

Beside the unified first-order line found within our model 
for YBCO \cite{Dietel3} (seen before 
within the Ginzburg-Landau model in Ref.~\onlinecite{Li2}), 
a third-order 
glass transition line emanates near the critical 
point on the melting line as a phase boundary between 
the Bragg-glass and the vortex-liquid 
 phase. We have shown in Ref. \onlinecite{Dietel3} by using 
hyperscaling relations 
that the higher-order character (more than second-order) 
of this line is in 
accordance with experiments and numerics 
which determines the scaling of the disorder phase correlation length 
\cite{Fisher1, Gammel1}. This glass transition line  
exists also  for BSCCO. 
But beside this parallel  of the glass transition lines of YBCO and BSCCO,  
Beidenkopf {\it et al.} \cite{Beidenkopf1} found for BSCCO 
an additional second-order glass transition 
line in the Bragg-glass phase by using also the 
magnetic shaking technique. The line showed up by 
plotting the derivate 
of the magnetic induction field with respect to the temperature. A jump 
was observed which also  exist for the glass transition line 
in the vortex-glass 
phases. Thus in contrast to YBCO, they found for BSCCO that  
both lines are of second-order characteristics.  
Both lines meet in a 
single point within experimental uncertainties. 
This point is not the critical point found for general 
doping \cite{Beidenkopf2} which is characterized by a 
vanishing of the entropy jump \cite{Avraham1} being the 
maximum of the unified 
first-order line. We label both  second-order lines 
in Figure 1 by $ T_d $. The intersection with the first-order line 
is denoted by GP2. 
Both lines divide the vortex-glass phase named by VG2 from a phase 
named VG1 in Fig.~1 lying in the high-magnetic-field part above the 
first-order line. In the low-magnetic-field part the $ T_d $ line divides 
two Bragg-glass phases denoted by BG2 and BG1 in Fig.~1   

Finally, we show in Fig.~1 a possible  additional phase boundary labeled by 
$ T_{x} $ which was found by Fuchs {\it et al.} \cite{Fuchs1} 
by measuring the vortex 
penetration through surface barriers. A similar line was also found 
by magnetization measurements \cite{Shibauchi1}. 
This line divides the vortex liquid denoted by VL from 
the phase VG1 shown in Fig.~1. It meets the melting line 
in a point to be referred as GP1.
Note that it is not experimentally clear whether the $ T_x $-line 
 has the characteristic of a phase transition. The  
$ T_{x} $-line  does not correspond to the irreversibility line where 
magnetization 
sweeps show hysteresis. 
The position of this line in the case of BSCCO 
is mostly influenced by surface barriers \cite{Raphael1} in contrast 
to YBCO where the pinning mechanism is responsible for the irreversibility. 
This leads to a coincidence of the irreversibility line with 
the glass transition line between the vortex-glass and vortex-liquid
\cite{Nishizaki1}. 
It was shown in Ref.~\cite{Shibauchi1} via Josephson plasma 
experiments that the $ T_x $-line is not accompanied by a 
Josephson decoupling between the layers ruling out the possibility  
of a transition from vortex lines to weakly coupled pancake vortices.
That such a transition exist was proposed theoretically in Refs.~
\onlinecite{Glazman1, Daemen1, Goldin1}.   
So far we  point out, that it is not experimentally clear yet  
what kind of phase VG1 is \cite{Fuchs1}.  
 There are, for example, hints that 
this phase could be a disordered flux line liquid 
\cite{Blasius1}. This is suggested by muon spin rotation experiments 
which, however are  
 in contradiction to other experiments which reported  
Bragg-peaks in this phase \cite{Forgan1}. In the interpretation 
of VG1 as a disordered flux line liquid, VG2 consists of  
a quasi-two dimensional vortex solid. 

It is the purpose of this paper
to  investigate the above phase transitions
in the  defect melting model 
mentioned earlier \cite{GFCM2, Dietel1, Dietel2} used in 
Ref.~\onlinecite{Dietel3} 
to calculate the phase diagram of YBCO. 
We will first review  briefly the model. A more comprehensive
discussion can be found in our former papers and the book \cite{GFCM2}.  
In contrast to YBCO, BSCCO is a strongly layered material where 
the coupling between the layers is described by the Josephson coupling 
in the Hamiltonian of the system. For high magnetic fields 
beyond the first-order line, one obtains a suppression of the  
Josephson coupling between  the layers \cite{Shibauchi1,Gaifullin1} 
with respect to the electromagnetic coupling. 
In our elastic model with defects, we can not get 
this decoupling. 
We shall accommodate it effectively via an appropriate  
modification of the elastic moduli of the vortex lattice system in 
this region of the magnetic field. With the help of the 
elastic moduli of Brandt \cite{Brandt1} for BSCCO  we show that one expects 
two thermal decouplings for the vortex strings in the 
liquid phase, corresponding to the two glass transition lines 
in Fig.~1, in contrast to the single line in YBCO 
\cite{Dietel3}. On this way, we carry out the M{\'e}zard-Parisi analysis 
for  the Hamiltonian  of the vortex lattice system with pinning. 
It consists of a variational approach to fit the free energy 
of the replicated system with the free energy of a quadratic Hamiltonian.
We obtain an almost vertical third-order depinning glass transition line 
in the liquid high-temperature phase located in the vicinity of the 
$ T_x $ line in Fig.~1 separating  
a full replica symmetric saddle point solution at high temperatures and  
a full replica symmetry broken  solution at lower temperatures.     
We show that the saddle point 
equation to the variational free energy has no solution for very low 
temperatures. This is also the case when going 
beyond the M{\'e}zard-Parisi 
theory within variational perturbation theory \cite{Kleinertpath1}.
This is a systematic extension of the  M{\'e}zard-Parisi theory to  
higher orders. 
It is well-known phenomenon of higher-order  variational perturbation  
expansion of the quantum mechanical anharmonic oscillator, 
that variations of the trial free energy  do not necessarily have  
to show a minimum or a maximum \cite{Kleinertpath1}, where 
the odd orders of variational perturbation theory possess a minimum,   
but even orders have no saddle points but only turning points. 
It is shown in Ref.~\cite{Jahnke1} for the 
anharmonic oscillator that also turning points are acceptable.
This is the principle of minimal sensitivity.   
Motivated by good results for the anharmonic oscillator  
we generalize the variational approach of  M{\'e}zard-Parisi
by using the principle 
of minimal sensitivity for the calculation of the variational free energy. 
With the help of this extension we shall obtain a variational 
free energy in the whole interesting regime of the $ H-T $  phase diagram
for the vortex lattice. 
This phase diagram looks rather similar to the phase diagram in Fig.~1.
The glass transition line $ T_d $ corresponds then within our model 
to the temperature where saddle point solutions to the variational 
free energy stops to 
exist and the best solution  
corresponds to turning points at lower temperatures where these turning point 
solutions are still full replica symmetry broken. 
The transition show second-order characteristic 
and can be interpreted as a thermal depinning transition where an almost 
equally displaced substring due to disorder forming a plateau decouples 
from the impurities due to temperature fluctuations.      

Finally, we point out here as was also the assumption 
for YBCO \cite{Dietel3} that we will 
only consider the phase diagram in the $ H-T $ regime 
in the vicinity of the 
melting line. Going beyond this restriction would take much more effort being 
out of the scope of this work.

The paper is organized as follows: 
In Section II and Section III 
we state the model and the M{\'e}zard-Parisi approach 
to the free energy of the vortex lattice system for BSCCO. 
In Section IV  we discuss solutions of 
the saddle point equation 
within the  M{\'e}zard-Parisi approach. In Section V we consider  
the regimes where this equation is not solvable.  
Section VI goes beyond lowest order variational perturbation theory 
using generalized principle of minimal sensitivity. 
Section VII discusses observable consequences of our theory. 
In the Appendices  A and B  we supply additional  material 
to Sections V and VI.

\section{Model}
The partition function to be used to describe 
the vortex lattice without disorder
was proposed in Ref.\,\onlinecite{Dietel2}. It is motivated by
similar melting models for two-dimensional square \cite{GFCM2}
and triangular \cite{Dietel1} crystals. In Ref.~\onlinecite{Dietel3} 
we derived from this partition function a low-temperature representation.
This corresponds to the 
partition function of the vortex-lattice in the crystalline phase   
\begin{equation}
Z_{\rm fl} = {\cal N} \prod_{{\bf x},i}\left[
 \int_{-\infty}^\infty\frac{u_i({\bf x})}{a} \right]
\exp\left[-\frac{1}{k_B T}
\left(H_{0}[u_i]+H_{\rm dis}[u_i]\right) \right]           \label{10}
\end{equation}
with the low-temperature Hamiltonian
\begin{align}
&  H_0[u_i]=
H_{\rm T \to 0}[u_i]  =  \frac{v}{2} \sum_{{\bf x}}
 (\overline{\nabla}_i u_i) (c_{11}-2c_{66}) (\overline{\nabla}_i u_i)
  \nonumber  \\
&   +\frac{1}{2}( \nabla_i u_j + \nabla_j u_i)\,  c_{66} \,
(\nabla_i u_j + \nabla_j u_i) +
  (\nabla_3 u_i) \, c_{44} \, (\nabla_3 u_i)     \nonumber \\
& = \frac{v}{2} \sum_{{\bf x}} (\nabla_i  u_L)\, c_{11} \,
(\nabla_i u_L) +
(\nabla_3 u_L) \, c_{44} \, (\nabla_3 u_L)  \nonumber \\
& + (\nabla_i  u_T) \, c_{66} \, (\nabla_i u_T) +
(\nabla_3 u_T) \, c_{44} \, (\nabla_3 u_T)  \,.
\label{15}
\end{align}  
Here $ {\bf u}_L = {\bf P}_L {\bf u} $ is the longitudinal part of the displacement 
where the projector $ {\bf P}_L $ is given by 
$ (P_L)_{jk} \equiv  - (1/\sqrt{|\nabla^2_i|}) \nabla_j \otimes
(1/\sqrt{|\nabla^2_i|}) \overline{\nabla}_k $. 
The transversal part of the displacements is then given by
$ {\bf u}_T = {\bf P}_T {\bf u} \equiv {\bf u}-{\bf u}_L $. 
By using the dual representation in the form of stress fields we 
obtain a high-temperature representation of the partition function. 
This partition function describes the vortex system in the fluid phase. 
We obtain a partition function of the form (\ref{10}) with Hamiltonian 
\begin{align}
 &  H_0[u_i]=
H_{\rm T \to \infty }[u_i]  =    \frac{v }{2} \sum_{{\bf x}}
 (\overline{\nabla}_i u_i) (c_{11}- c_{66}) (\overline{\nabla}_i u_i)
  \nonumber \\
&  \qquad \qquad +(\nabla_3 u_i) \, c_{44} \, (\nabla_3 u_i)  \nonumber \\
& =  \frac{v}{2} \sum_{{\bf x}}
 (\nabla_i  u_L)\, ( c_{11}- c_{66})
(\nabla_i u_L) +
(\nabla_3 u_L) \, c_{44} \, (\nabla_3 u_L)  \nonumber \\
& \qquad \qquad +
(\nabla_3 u_T) \, c_{44} \, (\nabla_3 u_T)                 \label{20}
\end{align}
and $ {\cal N}=  1/(4 \pi \beta)^N $.
In the following,  the subscripts $ i,j $ have 
values $ 1,2$,  and $l, m,n$ have values $  1,\ldots,3$ where $ N $ denotes 
the number of lattice sites.  
The parameter $ \beta $
 is
proportional to the inverse temperature,
 $ \beta \equiv v \,  c_{66}/k_B T (2\pi)^2  $,
where the volume $ v $
of the fundamental cell is equal to
  $ \sqrt{3} a^2 a_3 /2$ for the triangular lattice.
Here $a$ is the transverse distance of neighboring vortex lines,
and
 $ a_3 $ the persistence length
of the dislocation lines introduced in
 Ref.\,\onlinecite{Dietel2}. 
Note that $a_3$ is assumed to be
independent
on the disorder potential in the average \cite{Bem2}. 
Its value is given by \cite{Dietel2} 
\begin{equation} 
a_3 \approx  4 a \; \sqrt{\frac{2}{\pi}}
 \frac{  \lambda_{ab}}{\lambda_c} \,.  \label{22} 
\end{equation} 
The lattice derivates $ \nabla_i $ are built from the link differences 
around a plaquette in the triangular lattice. These expressions can be found 
in Refs.~\onlinecite{Dietel1, Dietel2}. By analogy $ \nabla_3 $ is the lattice 
derivate in z-direction.

The second term in the exponent
of (\ref{10})
\begin{equation}
 H_{\rm dis}[u_i]
= \sum_{{\bf x}} V({\bf x}+ {\bf u}),
 \label{25}
\end{equation}
accounts for disorder.
We have suppressed the spatial arguments
of the elastic parameters, which are
functional matrices
$ c_{ij}({\bf x},{\bf x}')\equiv
 c_{ij}({\bf x}-{\bf x}') $.
Their precise forms were first calculated by Brandt \cite{Brandt1}
and  generalized
in Ref.~\cite{Dietel2} by taking into account thermal softening relevant
for BSCCO.
The elastic moduli
$ c_{44} $ and $ c_{66} $
at low reduced magnetic fields $ B/H_{c2} <0.25 $ are given by 
\begin{eqnarray}
 c_{66} \! &  \!= \!&  \!\! \!\frac{B \phi_0}{(8 \pi \lambda_{ab})^2}
\,, \label{30} \\
 c_{44}\! & \! = \!&  \!\! \!
\frac{B^2}{4 \pi(1\!+ \! \lambda_{c}^2 k^2\! + \!\lambda_{ab}^2 k_3^2)}
\!+ \!
\frac{B \phi_0}{32 \pi^2 \lambda_{c}^2} \!
\ln  \! \frac{ 1 \! + \!
\frac{2 \lambda^2_{c}}{\langle u^2 \rangle} \! + \! \lambda_{ab}^2
 k^2_3}{
 1\! + \!   \lambda^2_{c}K_{\rm BZ}^2 \!   + \!\lambda_{ab}^2 k^2_3}
 \nonumber \\
& &  + \frac{B \phi_0}{ 32 \pi^2 \lambda_{ab}^4 k_3^2}
\ln \frac{1+ \lambda^2_{ab} k_3^2/(1+\lambda^2_{ab} K^2_{\rm BZ})}{
1+ \lambda^2_{ab} k_3^2/(1+2
 \lambda^2_{ab} /\langle u^2 \rangle)}
\, \label{35} 
\end{eqnarray}
where $ \lambda_c $ is the penetration depth in the $xy$-plane, 
and $ K_{\rm BZ} $ is the boundary of the
circular Brillouin zone $ K_{\rm BZ}^2=4 \pi B/\phi_0 $.
For BSCCO we use the two-fluid model \cite{Tinkham1}
   $ \lambda(T)=\lambda(0)[1-(T/T_c)^4]^{-1/2} $,
$ \xi(T)=\xi(0)[1-(T/T_c)^4]^{1/2}/[1-(T/T_c)^2]$ .
When calculating $ c_{44} $ in (\ref{30}) we have used
a momentum cutoff in the two-vortex interaction potential
$k\leq \sqrt{2}/\langle u ^2\rangle^{1/2}, $
rather than the inverse
of coherence  length $  1/\xi_{ab} $ as in Ref. 
\onlinecite{Brandt1}.
We shall not give here the explicit functional dependence of 
the elastic module $ c_{11} $. This can be found in 
Ref.~\onlinecite{Brandt1}. One can show that 
$ c_{11} \gg c_{44} $ and $  c_{66} $ in 
the vicinity of the melting line \cite{Dietel2}.  
This leads to the conclusion that one can neglect 
longitudinal fluctuations in comparison to 
transverse ones in the interesting  regime \cite{Dietel3}. 
This will be done in the following. 

The last term in $ c_{44} $ (\ref{35}) 
comes from the electromagnetic coupling 
between the layers. Its first two terms  are due to the Josephson 
coupling (both terms are vanishing for 
$ \lambda_c/\lambda_{ab} \to \infty $) resulting in a vanishing 
of these terms in the case of zero Josephson coupling. 
It is possible to find approximations for $ c_{44} $ in (\ref{35}) leading 
to tractable results for the calculation of the free energy expressions
of (\ref{10}). In Ref. \onlinecite{Dietel2} we used the approximations 
 \begin{equation}
  c_{44}(  k ,k_3 )  \approx \left\{ \large 
\begin{array}{c} 
\frac{B \phi_0}{32 \pi^2 \lambda_{ab}^2 (1+\lambda_{ab}^2 K^2_{\rm BZ})}
  \, \, ~{\rm for}~  k_3 \lesssim \frac{1}{\lambda_{ab}}\,,
    \\
\frac{B \phi_0 \ln(1\!+\!2B \lambda_{ab}^2/\phi_0 c_L^2) }{
 32 \pi^2 \lambda_{ab}^4 k_3^2}    ~{\rm for}~
k_3 \gtrsim \frac{1}{\lambda_{ab}}
\end{array} 
\right. \label{37} 
\end{equation} 
which are justified for $ |k_3| \lesssim \pi/a_3 $.  
Here,  the Lindemann parameter $ c_L^2=
\langle u^2 \rangle_{T \to 0} /a^2 $ restricted to the 
transversal fluctuations is given by 
\begin{eqnarray}
 c_L^2\! & = & \!\frac{a_3^2 }{a^2 v}  \frac{k_B T}{V_{\rm BZ}}
\int_{\rm BZ} \! \!d^2k dk_3 \frac{1}{c_{44}}  
\frac{1}{\frac{c_{66} a_3^2}{c_{44}}
  K^*_j K_j  + a_3^2 K^*_3 K_3 }   \label{39} 
\end{eqnarray}
 where the average is taken with respect to the
low-temperature Hamiltonian (\ref{15}) without disorder 
representing the elastic energy of
the vortex lattice. 
The momentum
integrations in (\ref{39}) run over the Brioullin zone
of the vortex lattice
whose volume is
$ V_{\rm BZ}= (2 \pi)^3 /v $, as indicated by the subscript BZ.
$ K_j $ is the Fourier transform of $ \nabla_j $ \cite{Dietel2}. 
Approximation (\ref{37}) for $ c_{44} $ is correct    
in the regime $ B \pi^3 \lambda_{ab}^2/ 8 \phi_0 \ln(1/c_L^2) 
\lesssim  1 $ which is valid 
in the vicinity of the melting line \cite{Dietel2}. In this regime we
obtain that $ c_{44}( k ,k_3 ) $ 
is dominated by the last term in (\ref{35}) for $ |k_3| <  \pi/a_3 $. 
For higher magnetic fields than the disorder induced first 
order BG2-VG2 line (see Fig.~1) we have  
$ B \pi^3 \lambda_{ab}^2/ 8 \phi_0 \ln(1/c_L^2)  \gtrsim  1 $ \cite{Dietel2}  
meaning that the first term in $ c_{44} $ (\ref{35}) 
is dominated over the third term in the region 
$ k_3 \approx \pi/a_3 $. 
This implies that the approximation 
(\ref{37}) would result in a wrong approximation 
for the magnetic field regime above the BG2-VG2  line. 
We can see from (\ref{15}) and (\ref{20}) that the 
string tension $ c_{44} $ is not renormalized going from the vortex 
lattice to the vortex liquid. For deriving the full elastic constants 
(\ref{30}), (\ref{35}) one uses a quadratic approximation for 
the Josephson coupling cosine phase difference term in  the Ginzburg-Landau 
model for BSCCO. 
It was shown in Refs.~\onlinecite{Koshelev1, Horovitz1} theoretically   
and in Refs.~\onlinecite{Shibauchi1, Gaifullin1}  
by determing Josephson plasma frequencies for BSCCO
that one gets a suppression of the full Josephson 
energy between the layers 
when going from the vortex solid to the vortex liquid 
crossing the BG2-VG2, BG1-VG1 line. 
This leads effectively to a softening of the Josephson 
terms of  $ c_{44} $ in the VG1 and  VG2  phases 
being the first two summands in (\ref{35}). 
This justifies to use (\ref{37}) as a good   
approximation for the full string tension in the 
whole interesting regime  when also including Josephson decoupling.

The disorder potential $ V({\bf x}) $ due to pinning is assumed
to possess the
Gaussian short-scale correlation function
\begin{eqnarray}
 \overline{V({\bf x}) V({\bf x}')} & = &  \Delta(x_i-x_i')
\delta_{x_3,x_3'}  \label{42} \\
 & = &
d(T)\,  a_3 \,  \frac{\phi_0^4 \,\xi^3_{ab}}{\lambda_{ab}^4}
K(x_i-x_i') \,  \delta_{x_3,x_3'}      \nonumber
\end{eqnarray}
where $ K(x_i-x_i') \approx 1/(\xi')^2  $ for
$ |{\bf x}-{\bf x}'|<\xi' $, and zero elsewhere.
The parameter $ \phi_0 $ is the magnetic flux quantum 
$ \phi_0=hc/2e $, and parameter
$ \xi' $ is the correlation length of the impurity potential, which
has a similar value as the coherence length $ \xi_{ab} $
in the $xy$-plane. 
In the following, we use an effective
disorder correlation function with the  Fourier transform
\begin{equation}
 {\hat K}(q ) = 2 \pi \exp(- {\xi'}^2 q_i^2/2 )
 \qquad  
  \label{44}
\end{equation}
leading also to an exponentially  vanishing of the disorder
correlation function  in real space.
In Ref.~\onlinecite{Dietel3} we have used this form for the correlation 
function in the solid phase for YBCO. In the present material 
BSCCO, this is even more justified because the disorder potential 
looks $\delta$-like for the vortices,  
due to the  large thermal fluctuations of the vortices near the melting 
transition line \cite{Dietel2}.

The temperature dependence of the parameter $ d(T) $ has two sources. 
One is the temperature dependence of the correlation length, the other 
is based on the
pinning mechanism where we discuss in the following the
$ \delta
T_c $-pinning or $ \delta_l $-pinning mechanisms \cite{Blatter1}
\begin {eqnarray}
 d(T)   & = &     d_0  (1-T/T_C)^{-1/2}  \quad  {\rm for}
\quad   \delta T_c-{\rm pinning}  \,,    \label{46}  \\
  d(T)  &  = &   d_0  (1-T/T_C)^{3/2}  \quad  \; \; {\rm for} \quad
\delta l-{\rm pinning}   .     \label{48}
\end{eqnarray}

\section{M{\'e}zard-Parisi Method}

We now carry out the calculation of the partition 
function (\ref{10}) which is still 
complicated due to disorder. In 
Ref.~\onlinecite{Dietel3} we have done this for  
YBCO by using a quadratic approximation in the disorder strength. 
This leads to a reentrant behaviour of the melting line
in the $ H-T $ plane  which did not agree with experimental results. 
By using the variational approach of M{\'e}zard-Parisi \cite{Mezard1} 
to go  beyond the quadratic approximation this reentrant behaviour
is disappeared, leading  to good results for the form of 
the melting line and agreement to the transition line 
between the Bragg-glass and vortex-glass. Here we use again the 
M{\'e}zard-Parisi theory to perform a similar calculation in the case 
of BSCCO. 
In order to go beyond second-order perturbation theory in the impurity
potential, we use first the well known replica trick \cite{Edwards1}. 
The M\'ezard-Parisi theory   consists in replacing
the non-quadratic part of this replicated Hamiltonian
as quadratic with a possible  mixing
of replica fields. By using the Bogoliubov variational principle we can find
the best  quadratic Hamiltonian so that its free energy named  
$ F_{\rm var} $  
is as close as possible to the actual free energy
of the system. 
This means that we have to search the minimum of 
\begin{equation}
 F_{\rm var}= F_{\rm trial} +
\langle H- H_{\rm trial} \rangle_{\rm trial}    \label{50}
\end{equation}
with the harmonic trial Hamiltonian
\begin{equation}
H_{\rm trial} = \frac{v}{2}\sum_{{\bf x},{\bf x}'}
\sum_{\alpha, \beta}  {\bf u}^{\alpha}({\bf x})
{\bf G}^{-1}_{\alpha \beta}
({\bf x}-{\bf x}')
{\bf u}^{\beta}({\bf x}') \,.  \label{52}
\end{equation}
Here $ \langle \cdot \rangle_{\rm trial} $ stands for the average with
respect of the Gibb's measure of the trial Hamiltonian
$ H_{\rm trial} $, while  $ H $ denotes the replicated Hamiltonian. 
The indices $ \alpha $, $ \beta $ denotes the replicas.  

In the general form, the search for an extremum is a complicated 
problem.
A strong simplification for this was founded 
by Parisi for random-spin systems where he 
suggest to deal with a trial Hamiltonian within some subalgebra known 
as the Parisi algebra. This restriction can be  motivated 
by physical arguments \cite{Dotsenko1}. 
It will be clear soon for the solid as well as the fluid phase
that the transverse part of $ {\bf G}_{\alpha \beta} $ can be chosen 
to have the form
\begin{equation}
G^{-1}_{\alpha \beta}=
G_0^{-1} \delta_{\alpha \beta }+ \sigma_{\alpha \beta}    
 \label{58}
\end{equation}
where $ G_0 $  is the transverse part of the Green function 
of  the Hamiltonian $ H_{0}[u_i] $ (\ref{15}) in the solid phase and 
(\ref{20}) in the fluid phase.    

Within the Parisi-algebra, the self-energy matrix 
$ \sigma_{\alpha \beta} $ depends effectively only on one parameter 
\cite{Mezard1} (see also App.~B).
In the general form it is allowed 
to be a continuous function $ \sigma(s) $ with $ 0 < s < 1 $ \cite{Mezard1}.  
Then the variational free energy has the form
\cite{Mezard1,Dietel3}
\begin{align}
& \Delta f_{\rm var}= \Delta F_{\rm var}/N  \equiv  
f_{\rm var}(B[\Delta]) - f_{\rm var}(0)      \label{54}      \\
 & =    \frac{k_B T }{2 } \int_0^1 ds \; \bigg[ \frac{1}{s^2}
\int_0^{\Delta(s)} \! \! \! \! \!
d \Delta  \;\Delta \frac{d}{d \Delta} g(\Delta)  +
{\cal D} \left(2 B[\Delta(s)] \right) \bigg] \,,  \nonumber \\
& f_{\rm var}(0)  = -k_B T \bigg( \frac{1}{N}
\ln {\cal N}
\label{60}   \\
&  + \frac{1}{2} \bigg\{\frac{1}{
 V_{\rm BZ}} \int_{\rm BZ}
 d^2 k dk_3 \; \ln \left[{\rm det} \, \left(
\frac{2 \pi k_B T}{v a^2}  G_0 \right)\right]
+   {\cal D}(0) \bigg\}\bigg)  
\nonumber
\end{align}
 where
\begin{equation}
g(\Delta) =  \frac{1}{ V_{\rm BZ}} \int_{\rm BZ}  d^2k dk_3
\left(G_0^{-1} +
\Delta  \right)^{-1}   \,.   \label{70}
\end{equation}
$ N $ is the number of lattice sites, and  
$ G_0^{-1} $ is given by 
\begin{eqnarray} 
G_{0}^{-1}({\bf k}, k_3) & = & \frac{c_{44}}{a_3^2}\Big[2-2 \cos(k_3 a_3)\Big]
\nonumber \\
& + & 
\frac{c_{66}}{a^2} \Big[4-\frac{4}{3} \sum^3_{l=1}
\cos({\bf k} {\bf e}_l a)\Big]  
\label{80} 
\end{eqnarray} 
in the solid low-temperature phase corresponding to (\ref{15}), and 
\begin{equation} 
G_{0}^{-1}({\bf k}, k_3)=\frac{c_{44}}{a_3^2}\Big[2-2 \cos(k_3 a_3)\Big]
\label{85} 
\end{equation}  
in the liquid  high-temperature phase corresponding to (\ref{20}).   
Here $ {\bf e}_l $ are the three unit link vectors around 
a plaquette in the triangular lattice.     
The gap function $ \Delta(s) $ and the self-energy function
$ \sigma(s) $ are  related by
\begin{equation}
\Delta(s) = \int_0^s  ds' s' \frac{d \sigma(s')}{d s'} \,.\label{90}
\end{equation}
$ B[\Delta(s)] $ is given by
 \begin{align}
& B[\Delta(s)]  =   \frac{ k_B T}{v} \frac{1}{s} g[\Delta(s)] -
\frac{k_B T}{v}
\int_s^1 ds'
\frac{1}{s'^2} g [\Delta(s')]                  \nonumber  \\
& \quad \quad   =   \frac{ k_B T}{v} g[\Delta(1)] -\frac{k_B T}{v}
\int_s^1 ds' \sigma'(s')g'[\Delta(s')] \,.    \label{100}  
\end{align}

In order to find a saddle point of $ F_{\rm var}/N $ we have
to take the derivative of (\ref{54}) with respect to $ \Delta(s) $.
This results in \cite{Korshunov1} 
\begin{equation}
\sigma(s) = -  2 \,  \frac{k_B T}{v} \;
{\cal D}' \left(2 B[\Delta(s)] \right)  
\label{120}
\end{equation}
where $ {\cal D}'(x) $ is the derivative $ (d/dx) {\cal D}(x) $. 
The disorder
function  $ {\cal D} $ is given by \cite{Dietel3}
\begin{eqnarray}
  {\cal D} (2 \langle u^2 \rangle)   & = & 
 d(T)
 \frac{a_3 }{(k_B T)^2 } \frac{ \phi_0^4 \,
\xi_{ab}^3 }{\lambda^4_{ab}}
\!  \int\!   \frac{d^2q}{ (2 \pi)^2}\,  \hat{K}(q) \,
   e^{-\frac{q^2}{2} \langle u^2 \rangle}  \nonumber \\
& = &   d(T)
 \frac{a_3 }{(k_B T)^2 } \frac{ \phi_0^4 \,
\xi_{ab}^3 }{\lambda^4_{ab}} 
\frac{1}{\xi'^2 + \langle u^2 \rangle}   \,.
                    \label{122}    
\end{eqnarray}  

In the following, we discuss solutions of  this equation in the
cases that  $ \sigma(s) $ does not break the replica symmetry,
possesses one-step replica symmetry breaking, or a continuous replica 
symmetry breaking.

In order to solve (\ref{120}), we first have to calculate $ g(\Delta) $
(\ref{70}) which we will denote by $ g^{T \to 0}(\Delta) $
with (\ref{80}) in the solid phase, and  by $ g^{T \to \infty} (\Delta) $
with (\ref{85}) in the fluid phase. 
We shall use the elastic constants $ c_{66} $ of Eq.~(\ref{30}) and 
the approximation (\ref{37}) for $ c_{44} $. In the liquid  case, 
the result is  
\begin{eqnarray} 
g^{T \to \infty} (\Delta) & \approx &   
\frac{1}{2} \frac{a^2_3}{c_{44}^{(1)}}  
\frac{1}{  \tilde{\Delta}^{1/2}(1+\tilde{\Delta}/4)^{1/2}}  
  \nonumber  \\
& & 
+ 1.38  \frac{a_3^2}{c_{44}^{(2)}} 
\frac{1}{(1+Z^{(0)}_l \tilde{\Delta}/2)}\,,  \label{125}
\end{eqnarray}          
and for the vortex solid  
\begin{eqnarray} 
g^{T \to 0}(\Delta) & \approx &   
 \frac{0.098  \pi \, a a_3  }{
\sqrt{c_{66} c^{(1)}_{44}}} - \frac{\sqrt{3}}{2} 
\frac{1}{4 \pi} \frac{a^2}{c_{66}}  
\, \tilde{\Delta}^{1/2} \nonumber  \\
& & 
+ 1.38  \frac{a_3^2}{c^{(2)}_{44}} 
\frac{1}{(1+Z^{(0)}_l \tilde{\Delta}/2)} \,, \label{130}
\end{eqnarray} 
where $ \tilde{\Delta} \equiv \Delta \, a_3^2/c^{(1)}_{44} $, 
and $ c^{(1)}_{44} $ denotes the function 
$ c_{44}({\bf k},k_3 \to 0) $ of Eq.~(\ref{37}) for 
$ k_3 \lesssim 1/\lambda_{ab} $,  and $ c^{(2)}_{44} $
denotes $ c_{44} $ of Eq.~(\ref{37}) in the region $ k_3 \gtrsim 1/\lambda_{ab} $ 
for $ k_3= 1/a_3 $, i.e.
\begin{eqnarray}
c_{44}^{(1)} & = &  
\frac{B \phi_0}{32 \pi^2 \lambda_{ab}^2 (1+\lambda_{ab}^2 K^2_{\rm BZ})} \,,
\label{135} \\
c_{44}^{(2)} & = &  
\frac{a_3^2 B \phi_0 \ln(1\!+\!2B \lambda_{ab}^2/\phi_0 c_L^2) }{
 32 \pi^2 \lambda_{ab}^4 }  \,.   \nonumber 
\end{eqnarray} 
For the derivation of (\ref{130}) we have used 
the approximation $ a_3^2 c_{66}/ a^2 c^{(2)}_{44}\ll 1 $ valid in the 
vicinity of the melting line.     
We used further the abbreviation
\begin{equation} 
Z^{(0)}_l=2 \cdot 1.38 \, \frac{ c^{(1)}_{44}}{c^{(2)}_{44}} \sim 
\left(\frac{\lambda_c}{a}\right)^2  \gg 1 \,. \label{140} 
\end{equation} 
In the solid phase the following abbreviations will be useful  
\begin{eqnarray} 
Z^{(0)}_s & = &  \frac{1.38}{0.098 \pi } \frac{a_3}{a}
\frac{\sqrt{c_{66}c^{(1)}_{44}}}{c^{(2)}_{44}} \sim 
\frac{\lambda_c }{a}  
\frac{\lambda_{ab}}{a}  \gg 1   ,     \label{145}  \\  
Z^{(1)}_s & = & \! \! 1.38^2  \frac{16 \pi }{\sqrt{3}}\frac{a_3^2}{a^2}
\left(\! \! \frac{\sqrt{c_{66} c^{(1)}_{44}}}{c^{(2)}_{44}}\! \! \right)^2 \! \! \! \sim
\! \! 
\left(\frac{\lambda_c}{a }\frac{\lambda_{ab}}{a }\right)^2 \! \! \gg \!  1 . 
 \label{150} 
\end{eqnarray}
Note that $ \lambda_{ab} \approx a $   
in the vicinity of the critical point on the melting line 
where the disorder is most influential the shape of this line. 
In this regime we obtain  
large numbers on the right hand sides of (\ref{140}), (\ref{145}), 
and (\ref{150}).   
In (\ref{125}) and (\ref{130}) the last terms have their origin in the 
integration (\ref{70}) over momenta 
$ 1/\lambda_{ab} \lesssim |k_3| \le \pi/a_3 $. The other terms  
come from the integration over small momenta. 
Expressions 
$ g^{T \to 0}(\Delta) $ (\ref{130}) and 
$ g^{T \to \infty} (\Delta) $ (\ref{125}) are not 
exact results of the integration in (\ref{70}). They are good 
approximations for $ g(\Delta) $ but also for 
$ g'(\Delta) $ and $ g''(\Delta) $
in the region 
$ \tilde{\Delta} \lesssim 1 $ in the fluid phase and 
$ \tilde{\Delta} \lesssim ((\lambda_{ab}/a)^{2} /Z_l^{(0)})^{2/3} $  
for the  solid.    
It will be seen below that these are the relevant regimes 
for $ F_{\rm var} $. 

We now define the quantity  
\begin{equation}
A =
\frac{4}{k_B T} \frac{ c^{(1)}_{44} 
a^2 \; {\xi'}^2 }
{a_3} \,.   \label{500}
\end{equation}
which will be useful below.  
Comparing $ g(\Delta) $ in (\ref{125}) and (\ref{130}) 
with the corresponding expressions for YBCO we obtain 
that only the last terms are different. 
The first term in (\ref{125}) 
leads in the case of YBCO to the decoupling scenario in the 
fluid high-temperature phase \cite{Blatter1}. This is the regime  
where temperature fluctuations starts to dominate over disorder fluctuations 
for the coherently pinned vortex line pieces 
given by $ {\cal D}(0) A \sim 1 $. 
The length of such line pieces are  given 
the Larkin length $ L_c $ where disorder fluctuations 
grow  to value $ \xi' $. 
  
One can now show by generalizing the calculation of the 
vortex fluctuations due to pinning and thermal  
fluctuations  for YBCO in Ref. \onlinecite{Bem1}  
that the additional last term in (\ref{125})
causes a new length scale beyond the Larkin length. 
At this length scale, the vortex fluctuations are 
approximately constant forming  a plateau.  
This length scales like $ L_T=a_3 (Z_{l}^{(0)}) $ 
for thermal  fluctuations, and like $ L_D=a_3 (Z_{l}^{(0)})^{2/3} $ 
for disorder fluctuations at low temperatures.  
Beyond these lengths, the displacement fluctuations starts to increase 
proportional to the cubic distance due to pinning, 
and proportional to the distance for thermal  fluctuations.
Both lengths, that of the coherently pinned vortex line pieces  but also 
the vortex substring on the plateau can decouple due to thermal  fluctuations.

Below we find two different depinning phase transition temperatures:
One takes place when the temperature 
fluctuations exceed the disorder fluctuations over the 
coherently pinned vortex line pieces where the 
the Larkin length fulfills $ L_c > L_T $. This leads to the well-known 
depinning temperature of the coherently pinned vortex substring given by 
$ {\cal D}(0) A \sim 1 $ corresponding to the 
third-order phase transition (\ref{1210}) below. 
The second depinning transition takes place when the temperature 
fluctuations exceed the disorder fluctuations over the plateau in the
 regime where  the Larkin length is given by $ L_c < L_D $. 
This leads to the depinning temperature  $ {\cal D}(0) A \sim Z_l^{(0)} $ 
corresponding to the second-order phase transition (\ref{1220}) given below.
Of course one can also see both depinning temperatures mentioned above  
in  the temperature dependency of the Larkin length $ L_c(T) $  
\cite{Kierfeld1},

\section{Solution of the M{\'e}zard-Parisi 
saddle point equations} 
In the following, we discuss the solutions of the M{\'e}zard-Parisi 
Eqs.~(\ref{90}), 
(\ref{100}) and (\ref{120}) in the liquid and the solid phase.

\subsection{Liquid  Phase}
In order to solve the   M{\'e}zard-Parisi  equations 
we transfer the analysis for YBCO of  
Ref.~\onlinecite{Dietel3} to BSCCO.  
Note that by neglecting the second term in (\ref{130}) 
we obtain a similar expression for $ g^{T \to \infty}(\Delta) $  
as for YBCO  \cite{Dietel3}. This leads to the following results: 
The stable solution for $ \tilde{\Delta}(s) $ is replica 
symmetric for $ {\cal D}(0) A  \le 2/\sqrt{3}  $  
and full replica symmetry broken in the case 
$ {\cal D}(0) A  > 2/\sqrt{3}  $.
That for example the one-step replica symmetry 
breaking solution is not stable can be seen from the following fact: 
The one-step self-energy function $ \tilde{\Delta}_1 $  
is given by  \cite{Dietel3}  
\begin{align}
& \frac{8}{(\frac{\sqrt{3}}{2} \tilde{\Delta}_1 A)^2}  
{\cal D }(0) \label{510} \\
& \times  \frac{ 
 \left( \sqrt{\tilde{\Delta}_1}+ 2  
 \left[ \log\left(1 + Z^{(0)}_l  \frac{\tilde{\Delta}_1}{2} \right) - 
\frac{Z^{(0)}_l   \frac{\tilde{\Delta}_1}{2} }{1+ Z^{(0)}_l   
 \frac{\tilde{\Delta}_1}{2}}\right]  \right) }{
 \left[ 1+ \frac{2}{\sqrt{3}} \frac{2}{A}
 \left( 
 \frac{1}{ 
  \sqrt{\tilde{\Delta}_1}} + 
 \frac{Z^{(0)}_l}{1+Z^{(0)}_l 
 \frac{\tilde{\Delta}_1}{2}} 
\right) \right]^3 } = 1 \,.\nonumber 
\end{align}   
The one-step symmetry breaking 
solution of the saddle point equation (\ref{120})   
is stable when the replicon eigenvalue 
\begin{align} 
& \lambda= 1-  \frac{8}{(\frac{\sqrt{3}}{2}\tilde{\Delta}_1 A)^2} 
{\cal D }(0) 
\label{515} \\
&\times  
 \frac{ 
  \left(\sqrt{\tilde{\Delta}_1}+ 4  
 \frac{ \left(Z^{(0)}_l \frac{\tilde{\Delta}_1}{2}\right)^2 } 
 {\left(1 + Z^{(0)}_l  \frac{\tilde{\Delta}_1}{2} \right)^2}\right)
 } 
{
 \left[ 1+ \frac{2}{\sqrt{3}} \frac{2}{A}
  \left( 
  \frac{1}{ 
   \sqrt{\tilde{\Delta}_1}} + 
  \frac{Z^{(0)}_l}{1+Z^{(0)}_l 
  \frac{\tilde{\Delta}_1}{2}}
  \right) \right]^3 
 }        \nonumber 
\end{align} 
is larger than zero. By comparing (\ref{510}) with (\ref{515}) we obtain that 
$  \lambda<0 $ when $  Z^{(0)}_l \tilde{\Delta}_1/2 \lesssim 13  $ meaning 
that the one-step replica symmetry breaking solution is unstable in this 
range.   
More generally one can show similarly  as in Ref. 
\onlinecite{Dietel3} that all finite 
step replica symmetry breaking solutions are unstable for 
$ {\cal D}(0) A  \gtrsim 2/\sqrt{3} $. 

Thus, we expect a continuous replica symmetry 
breaking solution in this parameter range. Note 
that continuous step replica symmetry 
breaking solutions of the saddle point equation (\ref{120}) 
are stable in general \cite{Carlucci1,Dietel3}. 
We now calculate this solution
by using the methods given in Ref.~\onlinecite{Dietel3}.    
First, the full replica symmetric solution 
for $ {\cal D}(0) A \le 2/ \sqrt{3} $ is given by  
\begin{equation}
\tilde{\Delta}(s) =0   \quad   {\rm for }  \quad  \frac{\sqrt{3}}{2} 
{\cal D}(0) A \le 1   \,.
 \label{522} 
\end{equation} 
The continuous replica symmetry broken solution 
for $ {\cal D}(0) A \gtrsim 2/ \sqrt{3}  $ can be derived  
from the saddle point equation (\ref{120}) by differentiating 
both sides with respect to  $ s $ resulting in 
\begin{equation}  \sigma'(s) = - \sigma'(s) 
4\left(\frac{k_B T}{v}\right)^2 g'[\Delta(s)] 
{\cal D}''\left(2 B[\Delta(s)] \right)          \,.          \label{525}  
\end{equation}  
This means that $ \sigma(s) $ is either constant or solves equation 
(\ref{525}) divided by $ \sigma'(s) $. Dividing (\ref{525}) by 
$ \sigma'(s) $ and forming the derivate with respect to $ s $, 
we obtain with (\ref{90}) 
\begin{align} 
& 2 \left(\frac{k_B T}{v}\right)  g'[\Delta(s)]^2 \; 
{\cal D}'''\left(2 B[\Delta(s)] \right)   \nonumber \\ 
& = -
s \, g''[\Delta(s)] {\cal D}''\left(2 B[\Delta(s)] \right) \,.    \label{528} 
\end{align} 
Equations (\ref{525}) and (\ref{528}) can be solved algebraically 
for the unknown functions $ \Delta(s) $ and $ B[\Delta(s)] $ leading to  
\begin{widetext} 
\begin{align}
 \tilde{\Delta}(s)=  
   \left\{ \begin{array}{c c c } 
0 &  {\rm for } & s \le \frac{1}{( \frac{\sqrt{3}}{2} {\cal D}(0) A )^{1/3}} 
\,, \\
\frac{\left(1+(\tilde{Z}_l^{(0)})^2\tilde{\Delta}^{3/2}\right)^{5/3} 
}{1+\frac{2}{3} 
(\tilde{Z}_l^{(0)})^3\tilde{\Delta}^{5/2}} = \left(\frac{\sqrt{3}}{2}
{\cal D}(0) A\right)^{1/3} 
s &  {\rm for } &  \frac{1}{(\frac{\sqrt{3}}{2} {\cal D}(0) A  )^{1/3}} 
 \le s \le s_c \,, \\
 \frac{\left(1+\tilde{Z}^{(0)}_l\tilde{\Delta}^{1/2}\right)^3 }{1+ 
(\tilde{Z}^{(0)}_l)^2\tilde{\Delta}^{3/2}} =  \left(\frac{\sqrt{3}}{2} 
{\cal D}(0) A 
\right)
& {\rm for } &  s_c \le s \le 1
\end{array} \right. \label{530} 
\end{align}
\end{widetext} 
where we used $ Z^{(0)}_l/A \approx c_L^2 a^2 / 2 \xi^2 \gg 1 $ 
in the vicinity of the melting line 
\cite{Dietel2} and the abbreviation 
$ \tilde{Z}_l^{(0)}\equiv Z_l^{(0)}/ (1+Z^{(0)}\tilde{\Delta}/2) $. 
This means that $ \tilde{\Delta} $ is constant for $ s \ge s_c $. 
The constant $ s_c $ is given by the equation  
\begin{equation} 
\frac{\left(1+(\tilde{Z}_l^{(0)})^2\tilde{\Delta}^{3/2}(s_c)
\right)^{5/3} }{1+\frac{2}{3} 
(\tilde{Z}_l^{(0)})^3\tilde{\Delta}^{5/2}(s_c)} = 
\left(\frac{\sqrt{3}}{2} {\cal D}(0) A\right)^{1/3} s_c  \,. \label{540}
\end{equation}   

Finally, we can calculate the disorder part of the 
variational free energy $ \Delta f_{\rm var} $ (\ref{54}) 
in the liquid phase \cite{Dietel3}. With 
$\Delta f_{\rm var} =  \Delta f^{\rm kin}_{\rm var} + 
\Delta f^{\rm pot}_{\rm var} $ where $ \Delta f^{\rm kin}_{\rm var} $ 
is the kinetic part represented by the first term in (\ref{54}) 
of the disorder energy and 
$ \Delta f^{\rm pot}_{\rm var} $ is the potential energy part 
of the disorder energy (second 
term in (\ref{54})), we obtain   
\begin{align}
& \Delta f^{\rm kin}_{\rm var}  =  -\frac{k_B T}{4} 
 \Bigg[\int\limits_{\frac{1}{(\frac{\sqrt{3}}{2}D(0)A)^{1/3}}}^{s_c}
\!\!\!\!\!\!\!\!\!   ds \frac{1}{s^2} \left( 
\tilde{\Delta}^{1/2}(s) + \frac{(\tilde{Z}_l^{(0)})^2}{4} 
  \tilde{\Delta}^{2}(s) \right)    \nonumber \\ 
&                -  \left(1-\frac{1}{s_c}\right) 
 \left( \tilde{\Delta}^{1/2}(s_c) +  
\frac{(\tilde{Z}_l^{(0)})^2}{4} 
\tilde{\Delta}^{2}(s_c) \right) \Bigg]    \label{550} 
\end{align}   
and 
\begin{align}
& \!\!\!\!\!\!\!\!\! \!\!\!\!\!\!\!\!\! \!\!\!\! 
\Delta f^{\rm pot}_{\rm var}  =  \frac{k_B T}{4} 
\left(\frac{\sqrt{3}}{2} {\cal D}(0) A\right)^{2/3} 
\nonumber \\ 
& \times  \Bigg[\int\limits_{\frac{1}{(\frac{\sqrt{3}}{2} D(0)A)^{1/3}}}^{s_c} 
\!\!\!\!\!\!\!\!\!   ds \frac{\tilde{\Delta}^{1/2}(s)}{
\left(1+(\tilde{Z}^{(0)}_l)^2 \tilde{\Delta}^{3/2}(s) \right)^{1/3}}  
     \nonumber \\ 
&                +  \left(1-s_c\right)
\frac{\tilde{\Delta}^{1/2}(s_c)}{
\left(1+(\tilde{Z}^{(0)}_l)^2 \tilde{\Delta}^{3/2}(s_c) \right)^{1/3}} 
  \Bigg]  \,. \label{560} 
\end{align}   
By taking into account (\ref{530}) and (\ref{540}) we obtain that 
the  glass transition line separating  the phases VG1-VL at 
$ {\cal D}(0) A = \sqrt{3}/2  $ is of 
third-order. We found the same order for the depinning 
transition in YBCO \cite{Dietel3}.  
\subsection{Solid  Phase}
In the solid phase one can show that finite-step replica symmetry  
breaking solutions are unstable \cite{Dietel3}. Similar as in the discussion 
of the fluid phase in the last subsection we obtain the following 
continuous replica symmetry broken solution of the saddle point equation 
(\ref{120}) 
\begin{widetext}
\begin{align}
& \tilde{\Delta}(s)= \! \!   \left\{ \begin{array}{c c c } 
\tilde{\Delta}^{2/3} \frac{\left(1+ 
\tilde{Z}_s^{(1)}\tilde{\Delta}^{1/2}\right)^{5/3}}
{1+ 2 \tilde{Z}_l^{(0)}
\tilde{Z}_s^{(1)}\tilde{\Delta}^{3/2}}
= \frac{(2 \pi)^{2/3}}{3 (\sqrt{3}/2)^{1/3} }
   ({\cal D}(0) A)^{1/3}  \left(\frac{c_{66} a^2_3}{c_{44}^{(1)} a^2} 
\right)^{2/3}  s 
&  {\rm for } &  0 
 \le s \le s_c \,, \\
\tilde{\Delta}^{1/2}  \frac{ 0.098^3 
\left(1+\tilde{Z}^{(0)}_s\right)^3}{1+ 
\tilde{Z}^{(1)}_s \tilde{\Delta}^{1/2}} =  \frac{3}{64 \pi^4} 
\left(\frac{c_{66} a^2_3}{c_{44}^{(1)} a^2} 
\right)^{1/2}
({\cal D}(0) A)
& {\rm for } &  s_c \le s \le 1
\end{array} \right.  \label{600}
\end{align}
\end{widetext} 
where we used $ Z^{(0)}_l / A \approx c_L^2 a^2 / 2 \xi'^2 \gg 1 $ 
in the vicinity of the melting line 
\cite{Dietel2} and the abbreviation 
$ \tilde{Z}^{(i)}_s\equiv Z^{(i)}_s/ (1+Z^{(0)}_l\tilde{\Delta}/2) $.  
The constant $ s_c $ is given by equation  
\begin{align} 
& \tilde{\Delta}^{2/3}(s_c) \frac{\left(1+ 
 \tilde{Z}_s^{(1)}\tilde{\Delta}^{1/2}(s_c)\right)^{5/3}}
{1+ 2 \tilde{Z}_l^{(0)}
\tilde{Z}_s^{(1)}\tilde{\Delta}^{1/2}(s_c)} \nonumber \\
& \quad = \frac{(2 \pi)^{2/3}}{3 (\sqrt{3}/2)^{1/3} }
   ({\cal D}(0) A)^{1/3}  \left(\frac{c_{66} a^2_3}{c_{44}^{(1)} a^2} 
\right)^{2/3}  s_c  \,.
  \label{610} 
\end{align}   

Finally, we can calculate the disorder part of the 
variational free energy $ \Delta f_{\rm var} $ (\ref{54}). We obtain 
for this energy in the solid phase  
\begin{align}
& \Delta f^{\rm kin}_{\rm var}  =  -\frac{\sqrt{3}}{2} \frac{k_B T}{16 \pi} 
\left(\frac{c_{44}^{(1)} a^2}{c_{66} a^2_3} \right) \nonumber \\
& \times  
\Bigg[\int\limits_{0}^{s_c}
\!\!    ds \frac{1}{s^2} \left( 
\frac{2}{3} \tilde{\Delta}^{3/2}(s) + \frac{\tilde{Z}^{(1)}_s}{2} 
  \tilde{\Delta}^{2}(s) \right)    \nonumber \\ 
&                -  \left(1-\frac{1}{s_c}\right) 
 \left( \frac{2}{3} \tilde{\Delta}^{3/2}(s_c) +  
\frac{\tilde{Z}_s^{(1)}}{2} 
\tilde{\Delta}^{2}(s_c) \right) \Bigg]  \label{620} 
\end{align}   
and 
\begin{align}
& \Delta f^{\rm pot}_{\rm var}  =  \frac{k_B T}{4} 
\left(\frac{\sqrt{3}}{2}\right)^{1/3} (2 \pi)^{1/3}
\left(\frac{c_{66} a^2_3}{c_{44}^{(1)} a^2}\right)^{1/3}   
 ( {\cal D}(0) A)^{2/3}      \nonumber \\
& \times  \Bigg[\int_{0}^{s_c} 
 \! \! ds \frac{\tilde{\Delta}^{1/6}(s)}{
\left(1+\tilde{Z}^{(1)}_s \tilde{\Delta}^{1/2}(s) \right)^{1/3}}  
     \nonumber \\ 
&                \qquad +  \left(1-s_c\right)
\frac{\tilde{\Delta}^{1/6}(s_c)}{
\left(1+\tilde{Z}^{(1)}_s \tilde{\Delta}^{1/2}(s_c) \right)^{1/3}} 
  \Bigg] \,. \label{630} 
\end{align}

\section{Existence and stability of 
saddle point solutions}
Trying to solve the implicit equation for $ \tilde{\Delta}(s_c) $ in the 
liquid phase (last line in (\ref{530})) and 
the solid phase (\ref{600}) we obtain that in both cases a solution 
is not existing for very large $ {\cal D}(0) A \gtrsim  
({\cal D}(0) A)_{\rm max}$
corresponding to low temperatures or large disorder strengths according to 
(\ref{122}), (\ref{500}). 

We obtain from Eq.~(\ref{530}) taken at $ s=s_c $ or directly from (\ref{525}) 
for $ ({\cal D}(0) A)_{\rm max}$  at low temperatures  
\begin{eqnarray}  
({\cal D}(0) A)_{\rm max} & \approx  &  \frac{2}{\sqrt{3}}  
Z^{(0)}_l \sim \left(\frac{\lambda_c}{a} 
\right)^2  \,,  \label{707}   \\
\tilde{\Delta}_{\rm max} & \approx & 
\frac{3^{2/5} }{(Z_l^{(0)})^{6/5}}  \,,
\label{708}  \\
(s_c)_{\rm max} & \approx  &  1 \label{709}     
\end{eqnarray} 
where $ \tilde{\Delta}=\tilde{\Delta}_{\rm max} $ and $ s_c=(s_c)_{\rm max} $ 
at $ {\cal D}(0) A=({\cal D}(0) A)_{\rm max} $.  
In the solid phase we have from (\ref{600})
\begin{eqnarray}
({\cal D}(0) A)_{\rm max} & \approx  &   
\frac{2}{\sqrt{3}}  Z^{(0)}_l \sim 
\left( \frac{\lambda_c}{a}\right)^2 \,, 
   \label{712}     \\
 \tilde{\Delta}_{\rm max} & \approx & \frac{1}{(Z_l^{(0)}Z_s^{(1)})^{2/3}} 
\,,    \label{713}   \\     
(s_c)_{\rm max} & \approx &  \frac{9}{5}  \,.   \label{714}   
\end{eqnarray}   
 The calculation was done by maximizing the left hand side 
of the implicit equation of (\ref{530}) and (\ref{600}) for 
$ s_c \le s \le 1 $ with respect to $ \tilde{\Delta} $. This then gives 
the maximal $ {\cal D}(0) A $ value given by $ ({\cal D}(0) A)_{\rm max}  $
where we still get a solution for both implicit equations. 
Summarizing we obtain that the continuous replica symmetry broken solutions
in the liquid as well as the solid phase stops to exist for 
$ ({\cal D}(0) A)_{\rm max} \approx 2  Z^{(0)}_l/ \sqrt{3} $. 
It was shown in the last paragraph that 
stable solutions of the saddle point equations 
are infinite replica symmetry broken (where we have the restriction to  
$ {\cal D}(0) A \ge 2/\sqrt{3} $ in the liquid phase). 
More generally we obtain that every saddle point solution 
of (\ref{120}) irrespective of its form 
is unstable for $ \sqrt{3} {\cal D}(0) A/2 \gtrsim   Z^{(0)}_l $ 
because the replicon eigenvalue \cite{Dietel3} 
\begin{equation} 
\lambda= 1 + 4 \left(\frac{k_B T }{v}\right)^2 
g'(\Delta(1)) \, {\cal D}''\left[2 \frac{k_B T}{v}  g(\Delta(1))\right] 
\label{720}
\end{equation} 
where $\Delta(1) $ is the self-energy function at $ s=1 $, 
is negative in this range.    

We point out that in the liquid phase similar to the continuous replica 
symmetry breaking  solution discussed above also the  
one-step replica symmetry breaking solution (\ref{510}) is no longer 
existing for $  \sqrt{3} {\cal D}(0) A/2 \gtrsim   4   Z^{(0)}_l $.
This can be seen by Taylor expanding  the left hand side  
of (\ref{510}) with respect to $ Z_l^{(0)} \tilde{\Delta}_1/2 $.
Note the difference in the prefactor of $ Z^{(0)}_l $ 
compared to (\ref{707}), (\ref{712}).  
This leads us to the more generally assumption that there is no saddle 
point solution of (\ref{120}) for $ \sqrt{3} {\cal D}(0) A/2  $ larger than   
$ \sim  Z^{(0)}_l $. This is proved in  
Appendix A where it is shown that this is true 
for every finite-step replica symmetry broken solution of (\ref{130})
by using results derived in Ref.~\onlinecite{Dietel3}.

We point out that $ {\cal D}(0) A \sim   Z^{(0)}_l  $ is in fact 
a relevant parameter region for the glass 
transition line because we expect that the critical point 
is around $ {\cal D}(0) A \sim Z^{(0)}_l $.  
Here we use (\ref{130}) with the fact that the quadratic approximation  
to the disorder energy at the peak should be approximately   
$ k_B T $ \cite{Dietel3}, i.e. 
$ {\cal D}(c_L^2 a^2) k_B T \sim k_B T  $.    
Note, as is shown in Fig.~1, 
the glass transition lines separating  the phases BG2-BG1, 
VG2-VG1 crosses the 
first-order line BG2-VG2, BG1-VG1 in the vicinity of the critical point 
for optimal doping.  

In the variational perturbation treatment of the 
anharmonic quantum mechanical oscillator we obtain a similar phenomenon.  
The even variational approximations to the free energy posses no extremum 
in the variational parameter \cite{Kleinertpath1}. 
Only the odd perturbative orders where the M{\'e}zad-Parisi theory 
belongs to the lowest order approximation within this perturbation theory,  
has a true minimum. In order to see whether we have a 
similar situation here, i.e. 
whether higher order variational approximations to the free energy posses 
a physical plausible extremum for $ \tilde{\Delta}(s) $, we will calculate 
in the following section higher order variational approximations to 
the free energy.

\section{Beyond lowest order 
variational perturbation theory} 
In this section, we will go beyond lowest order variational 
perturbation theory outlined in Section III. Starting from (\ref{50}) 
we can immediately write down the next beyond lowest orders of the free energy 
$ F $ within  variational perturbation theory \cite{Kleinertpath1}   
\begin{align}
&  F_{\rm var}= F_{\rm trial} +
\langle H- H_{\rm trial} \rangle_{\rm trial,c}
  \nonumber \\
& - \sum_{l=2}^{m} 
 \frac{1}{l!}
\frac{(-1)^l}{(k_B T)^{l-1}} \langle (H- H_{\rm trial})^l \rangle_{\rm trial,c}
   \label{1000}
\end{align}
$ \langle (H- H_{\rm trial})^l \rangle_{\rm trial,c} $ is the 
the averaging of  $ (H- H_{\rm trial})^l $ with respect 
to the trial Hamiltonian (\ref{52}) where we only take the connected 
expectation value part which means for example in second-order 
$ \langle (H- H_{\rm trial})^2 \rangle_{\rm trial,c} =
\langle (H- H_{\rm trial})^2 \rangle_{\rm trial}-
\langle (H- H_{\rm trial}) \rangle^2_{\rm trial} $.  
In order to calculate the free energy $ F_{\rm var} $
as in Section III  
within $ m$'th order variational perturbation theory 
we limit the $ l $-sum in (\ref{1000}) to $l=m $. 
$ F_{\rm var} $ corresponds to the exact 
free energy of the system for $ m \to \infty $ which means that 
$ F_{\rm var} $ does not depend on the choice of the 
trial Hamiltonian $ H_{\rm trial} $. The truncated 
sum depends on the choice of $ H_{\rm trial} $. Since the infinite sum 
is $ H_{\rm trial} $ independent, the best truncated sum should depend 
{\it minimally} 
 on $ H_{\rm trial} $. A first approximation would be in taking a 
saddle point of $ F_{\rm var} $ with respect to the trial Hamiltonian 
$ H_{\rm trial} $ leading to (\ref{120}) in the case $ m=1 $. 

To calculate $ F_{\rm var} $ beyond lowest order  
for a trial Hamiltonian $ H_{\rm trial} $ within the Parisi algebra is not 
an easy task. When going beyond lowest order we expect that the continuous 
replica symmetry breaking self-energy functions 
are still the relevant ones as was shown in Section IV via stability 
considerations for the M{\'e}zard-Parisi case corresponding to first order 
variational perturbation theory. 
We carry out the calculation of the free energy in Appendix B within 
second-order variational perturbation theory ($ m=2$).
We show that  for  
$ {\cal D}(0) A >   ({\cal D}(0) A)_{\rm max} \sim 
Z^{(0)}_l $  there exist also in this case 
 no continuous solutions of the saddle point equation 
in this second-order case. 
Thus, in contrast to 
the anharmonic oscillator where the variational perturbation 
theory leads to a solution of the saddle point equations for every 
odd order \cite{Kleinertpath1} a similar phenomenon is not existent 
in our case. Without explicit proof we now state the conjecture that 
this is true for every finite order within variational perturbation theory. 
 This means that there exist  
no saddle point of $ F_{\rm var} $ for large $ \sqrt{3} {\cal D}(0) A/2 
 \gtrsim Z_l^{(0)} $ which is a relevant physical regime outlined 
at the end of the last section. 
One way out of this dilemma is to continue the continuous replica symmetry 
broken solutions given in (\ref{530}) for the liquid and (\ref{600}) 
for the solid to the regime 
$  {\cal D}(0)A  >  ({\cal D}(0)A)_{\rm max} \approx  
2  Z^{(0)}_l / \sqrt{3} $  
by looking closer to  
the anharmonic oscillator problem solved in Ref.~\onlinecite{Kleinertpath1} 
via  variational perturbation theory. As mentioned above, 
for equal orders within the variational perturbation 
expansion which means $ m \in 
2 \mathbb{Z} $ in (\ref{1000}) one does not find a saddle point with respect 
to the trial harmonic Hamiltonian $ H_{\rm trial} $ being a quadratic 
potential in the anharmonic oscillator case. 
There it is shown that one gets good accordance with numerical 
solutions of the Schr\"odinger equation when interpreting the requirement 
of the {\it minimally} dependence of $ F_{\rm var} $  on $ H_{\rm trial} $ 
mentioned above by 
a   vanishing of the second-order derivation of $ F_{\rm var} $
with respect to $ H_{\rm trial} $. 
This is equivalent to the demand that the first 
order variation of $ F_{\rm var} $ on $ H_{\rm trial} $ is minimal. 

Transforming this general rule to our case 
by using (\ref{120}), the self-energy function  $ \sigma(s) $ is given by 
the minimum of the functional   
\begin{equation}
{\rm Min}_{\sigma(s)}\left[\left|1 +  \frac{2}{\sigma(s)} \frac{k_B T}{v} \;
{\cal D}' \left(2 B[\Delta(s)] \right)\right|  \right] 
\label{1030}
\end{equation} 
where we assume as a first approximation that this minimum 
is not dependent on $s $. This leads us to the following result for 
$ {\cal D}(0) A \ge ({\cal D}(0) A)_{\rm max} $: \\ 
The solutions $ \tilde{\Delta}(s) $ of (\ref{1030}) where $ \sigma(s) $ and 
$ \Delta(s) $ related by (\ref{90}) are given by (\ref{530}), (\ref{600}) 
with the substitution $ {\cal D}(0) A \rightarrow ({\cal D}(0) A)_{\rm max} $. 
The variational energies $ \Delta f_{\rm var} $ are given by 
(\ref{550}), (\ref{560}) in the liquid phase and (\ref{620}), (\ref{630}) in 
the solid phase with the same  substitution. Furthermore, one has 
to multiply the potential part of the disorder energies (\ref{560}) and 
(\ref{630}) by a correction factor 
$ ({\cal D}(0) A)/({\cal D}(0) A)_{\rm max} $ for 
${\cal D}(0) A > ({\cal D}(0) A)_{\rm max} $.

Summarizing, we obtain for BSCCO 
a third-order glass transition in the liquid phase having its reason in the 
breaking of the full replica symmetry across the transition line 
at $ {\cal D}(0) A =2/\sqrt{3}  $. A similar transition was also found for 
YBCO \cite{Dietel3}. Beside this transition    
we will show in the next section that (\ref{1030}) leads additionally 
to a second-order 
glass transition line at $ {\cal D}(0) A\approx 2 Z^{(0)}_l/\sqrt{3} $  
in both phases. We point out that this transition is not reasoned in 
the generalization of the saddle point criterium for the variational 
free energy to the more general principle of minimal sensitivity (\ref{1030}). Up to now, we have only searched a saddle point of the 
variational free energy in the 
self-energy matrices  $ \sigma_{\alpha \beta}  $ of the Parisi form (see the 
discussion below (\ref{52})) which could be 
motivated physically \cite{Dotsenko1} as an Ansatz 
for the glassy-state self-energy matrices. Nevertheless, it 
could also be possible that the restriction to this subspace is the 
reason that we do not find a saddle point of the variational 
free energy for $ {\cal D}(0) A \gtrsim  2 Z^{(0)}_l/\sqrt{3}  $.
On the other hand it is clear that also in this case the leaving  
of the stable saddle point solutions from the subspace of self-energy
matrices of the Parisi form leads in general to a non-analytically 
of the free energy at the point 
$ {\cal D}(0) A  \approx 2 Z^{(0)}_l/\sqrt{3} $ 
and thus to a phase transition.

As we explained above the reason 
that the saddle point solutions of the variational free energy 
stops to exist within the M{\'e}zard-Paris theory 
lies in the non-solvability of (\ref{525}) for $ s=s_c $. 
This follows further from the fact that $ g(\Delta) \sim 1/\tilde{\Delta}$ 
for large $ \tilde{\Delta} $ and that
$ {\cal D}''(2 B[\Delta(s_c)]) \approx {\cal D}''(0) \xi'^6/B[\Delta(s_c)]^3$
for the relevant $ \Delta(s_c) $ values where  $ g $ 
begin to show the behaviour  
$ g(\Delta(s_c)) \sim 1/\tilde{\Delta}(s_c)$. In deriving the approximation 
for $ {\cal D}'' $ above we use $c_L^2 a^2 \gg \xi'^2 $ for BSCCO
(see the notes below (\ref{530}) and (\ref{600})).
In contrast to this we find for YBCO 
$c_L^2 a^2 \ll \xi'^2 $ leading to the existence of the  
saddle point solutions of the variational free energy 
in the whole $ H-T $ plane although  
we have also $ g(\Delta) \sim 1/\tilde{\Delta}$ for large 
$ \tilde{\Delta} $ in this case \cite{Dietel3}. This is the reason that 
one does not find the second-order glass transition line in YBCO.  

Finally, we note that Giamarchi {\it et al.} 
in Ref.~\cite{Giamarchi1} only consider the small $ \tilde{\Delta} $ behaviour 
of $ g(\Delta) $ which is presumably 
the reason that they did not find the second-order 
glass transition line at least in the solid phase. The reason that 
we can compare only our low-temperature solid phase results with results 
in this paper lies in the fact that they did not consider 
defects  as we do here being 
relevant in the high-temperature liquid phase. 
Note that they did not use temperature 
softened elastic constants in their calculation 
relevant for BSCCO \cite{Dietel2}.

\section{Observable Consequences}
In the following, we use the intersection criterium 
\cite{Dietel2} with variational free energies (\ref{54}) and 
(\ref{60}) to get the first-order line separating the 
phases BG2-VG2, BG1-VG1, BG1-VL. 
This results in 
\begin{align}
 & 
   B_m(T)  \approx  \frac{1}{192} \frac{1}{\sqrt{3} \pi^7}
 \frac{(1-(T/T_c)^4)^2}{\lambda_{ac}^2(0)   \lambda_{c}^2(0)}
 \frac{\phi_0^5}{(k_BT)^2}   \nonumber \\
&   
 \quad \times \exp\left[-\frac{2}{k_B T}
  \left(\Delta f^{T \to 0}_{\rm var }- \Delta f^{T \to \infty}_{\rm var}
\right) \right]               \label{1200}
\end{align}  
where $ \Delta f^{T \to \infty}_{\rm var } $ is given by 
the disorder part of the variational free energy which is the sum of 
(\ref{550}) and 
(\ref{560}) in the liquid case. 
$ \Delta f^{T \to 0}_{\rm var } $ corresponds to the disorder 
part of the variational free energy in the solid case given by (\ref{620}) 
and (\ref{630}). 

Beside this first-order transition line we obtained a third-order 
glass transition line of the depinning form in the fluid phase 
separating the VL and VG1 phases 
\begin{equation}
{\cal D}(0) A = \frac{2}{\sqrt{3}}         \label{1210} 
\end{equation}  
and a second-order glass transition line separating 
 BG1 with BG2 in the solid phase and 
VG1 with VG2 in the liquid phase given by (\ref{707}), (\ref{712}) 
\begin{equation} 
{\cal D}(0) A= ({\cal D}(0) A)_{\rm max} \approx  \frac{2}{\sqrt{3}}
Z_l^{(0)}    \,.  \label{1220} 
\end{equation} 
This means that we obtain within our analytical 
approximation a unified glass transition 
line in both phases in correspondence to the experimental findings 
shown in Figure 1.   

In the following figures, we use parameter values for optimal doped
BSCCO given by $ \lambda_{ab}(0) \approx 2300 $\AA, $ \xi_{ab}(0) \approx 
30 $\AA,
CuO$_2$ double layer spacing $ a_s=14$\AA, $T_c=90 K$ and
the anisotropy parameter $ \gamma = \lambda_c/\lambda_{ab} \approx 250 $.
Due to the small coupling between the layers, 
the Josephson form of the interlayer coupling leads 
to a non-negligible softening of $ \lambda_c $ or 
$ \gamma= \lambda_c/\lambda_{ab} $, respectively,   
as a function of $ B $ and $ T $.  In  Ref.~\onlinecite{Gaifullin1} it was 
found by Josephson plasma experiments
 that $ \lambda_c(B,T) $ is nearly of the form 
$  1/\lambda^2_c(B,T) \approx  (1+ F(B/B_m))/\lambda^2_c(0,T) $ 
with  some function $ F $ which can be found in \cite{Gaifullin1} and 
further that $ \lambda_c^2(0,T)/\lambda_c^2(B_m,T) \approx 0.6 $. 
This leads to 
$ \gamma =  \lambda_c(B,T)/ \lambda_{ab}  \approx 250 $ in the vicinity 
of the first-order line separating the phases BG2-VG2, BG1-VG1, BG1-VL. 
Here we used $ \lambda_c(0,T) / \lambda_{ab}(T) \approx 200 $
 as in Ref.~\onlinecite{Dietel2}.    

In Figure 2 we show (\ref{1200}) corresponding to the first 
order line separating phases BG2-VG2, BG1-VG1,  BG1-VL  
for $ \delta T_c $ pinning given by the correlation function (\ref{46}) 
(upper figure) and 
$ \delta l $ pinning (\ref{48}) (lower figure) for various 
constants $ d_0 $.       
The square points in the figure denotes the experimentally determined
first-order BG2-VG2, BG1-VG1, BG1-VL  line of Beidenkopf {\it et al.} in 
Ref.~\onlinecite{Beidenkopf1}. 
The $ d_0 $ values of the 
the straight (black) curves are  chosen in such a way, that we 
reproduce in one of the best ways the experimentally 
given curves of Beidenkopf {\it et al.} and also the glass 
intersection point GP2. We obtain $ d_0= 2.5 \cdot 10^{-6} $ in the 
$ \delta T_c $ pinning case and $ d_0= 4 \cdot 10^{-6} $ for $ \delta_l $ 
pinning. The curves of representative variations of these almost best 
parameter values are given by the (red) dotted lines in Fig.~2.    
We obtain discrepancies in the form 
of the first-order BG2-VG2, BG1-VG1, BG1-VL line from the experiment. 
There are a  large variety of the concrete 
forms of this  line in the literature (see for example 
\cite{Avraham1} for an almost horizontal BG2-VG2 line with a small kink 
near the intersection point GP2). The reason for the discrepancies 
 comes mainly from the sensitivity of the curve on the disorder function 
\cite{Dietel3} but also the neglection of the layerdness of BSCCO 
in our case could be one factor. 
Without taking into account 
dislocations, we expect a Josephson decoupling transition 
which is nearly temperature independent for low temperatures 
\cite{Horovitz1}. The melting line and the decoupling line 
lies on top of each other when taking into account dislocations 
leading to the first-order BG2-VG2, BG1-VG1, BG1-VL transition line. 
Note that the Josephson  decoupling is not complete over the 
transition line and further that the  latent heat due to the 
Josephson degree of freedom is only $16$ \% of the total 
latent heat over the first-order transition line \cite{Gaifullin1}.
The competition between the temperature independent 
decoupling transition and the temperature dependent 
three dimensional first-order line should take into account 
the correct form of the whole first-order line for layered materials.  

The small vertical marks on the curves in Fig.~2 denotes 
the glass intersection point GP2. 
We obtain especially for the $ \delta T_c $ pinning case 
differences in the location of the glass intersection point GP2 with the 
experiment. In all shown three $ \delta T_c $ pinning  
cases the glass transition point GP2 lies in the vicinity 
of the critical temperature $ T_c $ where in both pinning mechanisms 
also the glass intersection point GP1 is located. Summarizing, 
we obtain as was also the case for YBCO \cite{Dietel3} that the 
$ \delta l $ pinning mechanism gives a better accordance to the 
experimental curves and glass intersection points than the 
$ \delta T_c $ pinning mechanism.  
   
\begin{figure}[t]
\begin{center}
\includegraphics[height=9cm,width=8.5cm]{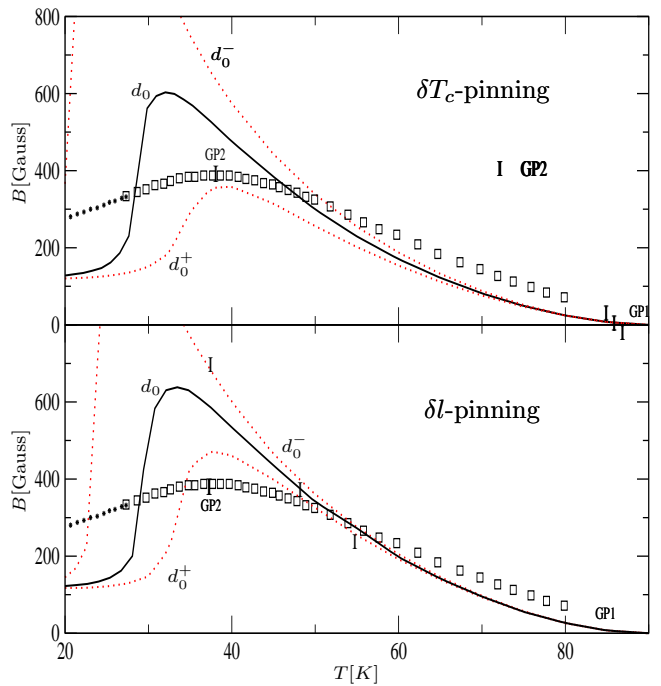}
\end{center}
 \caption{The BG2-VG2, BG1-VG1, BG1-VL first-order transition lines
$B_m(T) $ given in (\ref{1200}) for $ \delta T_c $-pinning (upper figure)
and  $ \delta l $-pinning  (lower figure). The solid (black) 
lines are calculated
with parameters for $ d_0 $ which gives one of the best
fits to the experimentally determined phase diagram  
in Ref.~\onlinecite{Beidenkopf1}
(square points) within the
pinning mechanism ($ d_0 =2.5 \cdot 10^{-6} $ for
$ \delta T_c $-pinning,
$ d_0 =4 \cdot 10^{-6} $
for $ \delta l $-pinning).
Dotted (red) curves are calculated by a variation of
these parameters given by $ d^\pm_0=(1 \pm 1/2) d_0 $.
The vertical markers denote the
intersection points of the VG2-VG1, BG2-BG1 glass transition line and the
disorder induced first-order line BG2-VG2, BG1-VG1 named GP2. GP1 denotes 
the intersection point of the third-order VG1-VL glass transition line 
with the first-order melting line BG1-VG1, BG1-VL.}
\vspace*{0cm}
\end{figure}

\begin{figure}[t]
\begin{center}
\includegraphics[height=6cm,width=8.5cm]{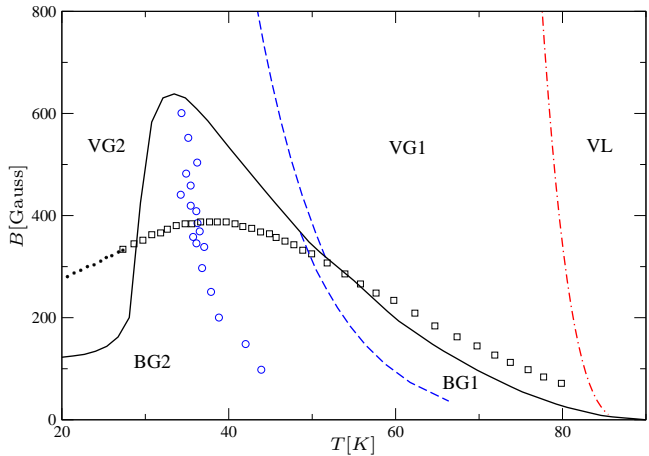}
\end{center}
 \caption{
 Phase diagram for BSCCO in lowest order variational 
perturbation theory corresponding to the M{\'e}zard-Parisi theory. 
Lines represent the
theoretical determined phase transitions between the various phases
 calculated for $ \delta l $ pinning with
 $ d_0 =4 \cdot 10^{-6} $   corresponding
 the solid line in the lower picture in Fig.~2.
Points represent the 
 experimentally determined phase diagram of Beidenkopf  {\it et al.}
 \cite{Beidenkopf1}.
The solid (black) line denotes the first-order BG2-VG2, BG1-VG1, BG1-VL
line calculated by (\ref{1200}).  
The glass transition lines BG2-BG1, VG2-VG1 are given by the (blue) 
dashed lines calculated from the numerical 
generalization to the approximation (\ref{1220}) as described 
in Section V. The red dashed-dotted line is the glass transition line VG1-VL 
derived by using expression (\ref{1210})}
\vspace*{0cm}
\end{figure}

\begin{figure}[t]
\begin{center}
\includegraphics[height=8.0cm,width=8.5cm]{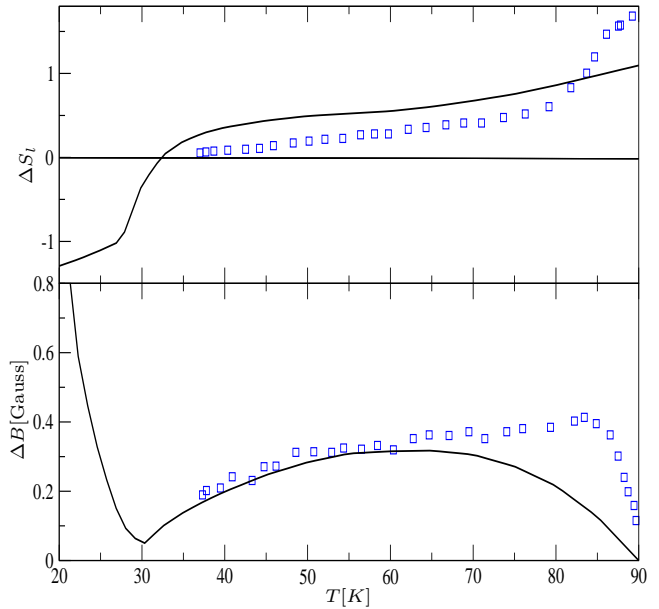}
\end{center}
 \hspace*{0.5cm}
 \caption{
Entropy jump $ \Delta S_l $  per double layer and vortex 
(upper figure), 
and the magnetic induction jumps
$ \Delta B $ (lower figure) 
over the first-order line BG2-VG2, BG1-VG1, BG1-VL. Calculations are based   
on expressions derived in Ref.~\onlinecite{Dietel3}.
In both figures we used  for the (black) solid lines 
the $ \delta l $-pinning mechanism  with
$ d_0 =4 \cdot 10^{-6} $  corresponding to the parameters of the phase 
diagram in  Fig~3.  
The points in both figures represent experimental values 
determined by Zeldov {\it et al.} \cite{Zeldov1}. } 
 \hspace*{0.0cm}
\end{figure}
In Figure 3 we show for $ d_0= 4 \cdot 10^{-6} $ 
with $ \delta l $ pinning correlated impurities the whole phase diagram 
calculated with (\ref{1200}), (\ref{1210}) and (\ref{1220}) 
corresponding to the parameter values $ d_0 $ of the (black) 
solid line in the lower picture in Figure 2. 
Note that $ H_{c2} $ can not be resolved in this figure 
being almost vertical directed on the right boundary. 
Again we show for comparison the 
experimentally determined phase diagram of Beidenkopf {\it et al.}
where the square points denote the first-order 
BG2-VG2, BG1-VG1, BG1-VL transition line. The (blue) circle points denote 
the experimentally determined $ T_d $ line BG2-BG1, VG2-VG1 of 
Beidenkopf {\it et al.}. This line has to be compared with the (blue) 
dashed lines VG2-VG1, BG2-BG1, calculated with (\ref{1220}) where we get small 
discrepancies in the intersection point on the first-order line 
of the upper high magnetic curve VG2-VG1 in the liquid phase with the small
magnetic field transition line BG2-BG1 in the solid phase. The reason is
that we did not use the analytical approximation 
$ \sqrt{3} ({\cal D}(0) A)_{\rm max}/2  \approx Z^{(0)}_l$ 
for  $ ({\cal D}(0) A)_{\rm max} $ valid in both phases 
but the numerical determined 
values calculated from the condition that 
(\ref{525}) stops to be solvable as described in Section V. 
As mentioned by Beidenkopf {\it et al.} in Ref.~\cite{Beidenkopf1} 
it could be experimentally possible that both lines do not intersect.
From Fig.~3 we obtain that the point GP2 does not coincidence with the maximum 
of the theoretical determined first-order BG2-VG2, BG1-VG1, BG1-VL transition 
line which coincides with the critical point \cite{Dietel3} 
(see also Fig.~4). This is 
possible for general doping \cite{Beidenkopf2}.  
Nevertheless, we  obtain a  discrepancy between the position of our   
glass transition lines and the 
experimental findings. One reason comes 
from the approximations to the elastic moduli carried out in 
Section II but also corrections to (\ref{125}), (\ref{130}) 
where we used $ a_3 c_{66} /  a^2 c_{44}^{(2)} \ll 1 $.  
These  approximations getting 
worse for higher magnetic fields \cite{Dietel2}.  
This leads to an additional bending of the 
first-order line in the direction to the temperature axis shown in 
Fig.~1 of Ref.~\onlinecite{Dietel2} without pinning.  
 To get the same effective bending of this line as in the experiments 
we have to use a smaller 
$ d_0 $ value leading to   BG2-BG1, VG2-VG1 lines 
located  at smaller temperatures according to (\ref{1220}).
Furthermore, a source of the additional bending can be also due to 
the decoupling transition between the Josephson layers 
as discussed above.   

Beside these reasons also the restriction to the lowest order variational 
perturbation approximation could be a source for the 
difference of our theoretical finding of the glass transition
line and the  experimental ones.
The calculation of the free energy within second-order variational 
perturbation theory is outlined in Appendix B. We did not carry out  
the calculation of the phase diagram within this order  
which is rather non-trivial being out of the scope of this work.  
 
Finally, the (red) dashed-dotted  line 
in Figure 3 shows the VG1-VL glass transition line calculated by the 
help of the depinning temperature formula (\ref{1210}). 
We do not show for comparison 
the $ T_x $ line of Fuchs {\it et al.} \cite{Fuchs1}
in the figure because they did not use an optimal doped crystal 
in the experiment.

In Fig.~4 we show in the upper picture the entropy jumps 
per double layer and vortex $ \Delta S_l $ and in 
the lower picture the 
magnetic induction jumps $ \Delta B $ over the first-order 
BG2-VG2, BG1-VG1, BG1-VL transition line. The (black) full line is calculated 
with $ d_0= 4 \cdot 10^{-6} $ in the $ \delta l $ pinning case 
corresponding to parameter values of the phase diagram in Fig.~3.
We used formulas derived in Ref.~\onlinecite{Dietel3} for the calculation. 
Note further, as was also the case in Ref.~\cite{Dietel2}, 
that we did not use corrections for $ \Delta B $ by considering 
explicitly the difference of the induction field $ B $ and 
the external magnetic field $ H$. These differences are negligible in the 
interesting  regime \cite{Zeldov1}. 
The square points (blue) are experimentally determined 
values measured by Zeldov {\it et al.} \cite{Zeldov1} 
for optimal doped BSCCO crystals. 
We note that there are other experiments in the literature for 
non-optimal doped crystals where $ \Delta S $ and $ \Delta B $ varies 
significantly \cite{Kadowaki1, Avraham1}. The reason for this difference is 
not clear. The largest difference in Figure~4  
between experimentally and theoretically determined curves is at
 high-temperatures near $ T_c $. As noted in \cite{Dietel2} this  comes mainly 
from  contributions of thermally activated vortex loops not inherent 
in our vortex lattice picture.  
 
In the paper of Beidenkopf {\it et al.} \cite{Beidenkopf1} the order  of the 
glass transition lines VG2-VG1, BG2-BG1 was determined by 
measuring the magnetic  induction field and its derivate with respect 
to the temperature across  this line. 
They found a jump of $ \partial B(H,T) / \partial T $ across the line
leading to the conclusion that this transition is of second-order. 
They also deduced from their experiment that the jumps over the 
glass transition 
line are of almost the same magnitude in the BG1-BG2 phase and the VG1-VG2 
phase \cite{Beidenkopf1, Beidenkopf2}. 
Nevertheless the displayed curves in their paper show 
a much smoother behaviour of the magnetic induction curve and its temperature 
derivate near the glass transition line in the BG2-BG1 
solid phase than in the VG2-VG1 phase.
The problem of determing the 
order of the transition comes mainly from a large noise on the 
magnetic induction curves having its reason presumably in the 
spatial and temporal inhomogeneities of the system.
This is the reason that Beidenkopf {\it et al.} 
in Ref.~\cite{Beidenkopf1} did not get a clear 
jump in the derivative in all measurements (while
the induction itself bends sharply) to allow a systematic quantitative study
of it \cite{Beidenkopf3}.

The magnetic induction field $ B $ is given by 
\begin{eqnarray}
B & = &  H+ \left[4 \pi (k_B T )\frac{\partial}{\partial  B} \frac{1}{N v} 
\ln (Z_{\rm nfl})+B \right]   \nonumber \\
 & &  
+ 4 \pi (k_B T) \frac{\partial}{\partial  B} \frac{1}{N v} \ln(Z_{\rm fl})   
  \label{1250}
\end{eqnarray}
where $ Z_{\rm nfl} $ is the partition function 
of the static non-fluctuating part of the vortex lattice.  
 In Fig.~5 we show the disorder part of the magnetic induction field 
(solid (black) curves) 
 $ B $ given by  
\begin{equation} 
 B_{\rm dis} = - 4 \pi \frac{\partial}{\partial  B}   
\frac{1}{v} \Delta f_{\rm var} \,,                \label{1260}
\end{equation}     
and also its  derivate with respect to $ T $
 (dashed blue curves), i.e. $  
\partial B_{\rm dis}/\partial T $ 
as a function of temperature for 
two different magnetic fields. As mentioned above, the 
magnetic induction contribution from the non-fluctuating part of the partition 
function   $ \log(Z_{\rm nfl}) $ is negligible in comparison to $ H $ in 
(\ref{1250}) to the induction field $ B $. This means that we obtain 
for the jump values over the glass transition line 
$ \Delta \partial B (H,T)/ \partial T 
\approx \Delta \partial B_{\rm dis}(B,T) / \partial T $.  

\begin{figure}[t]
\begin{center}
\includegraphics[height=12.0cm,width=8.5cm]{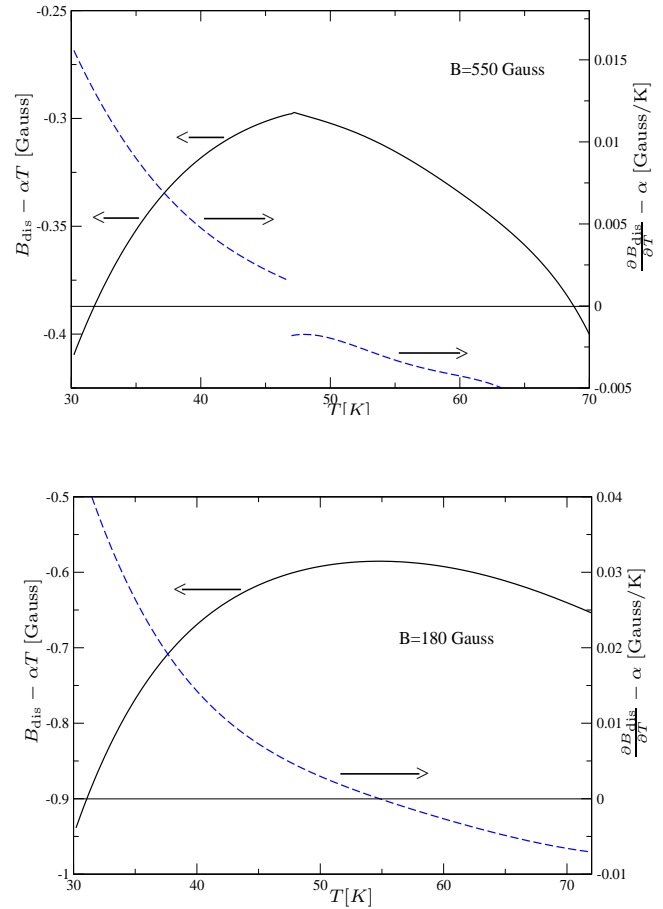}
\end{center}
 \caption{
The disorder part $ B_{\rm dis} - \alpha T $ 
of the magnetic induction field  defined by (\ref{1260}) 
(solid (black) curves), and its derivate with respect to 
$ T $, i.e.  $ \partial B_{\rm dis}/\partial T  - 
\alpha  $ (dashed (blue) curves) for two 
different magnetic fields $ B $ either in the VG2-VG1 phase (upper figure) 
or in the BG2-BG1 phase (lower figure). Here $ \alpha $ is some subtraction 
parameter determined such that the 
$ B_{\rm dis} $ curve is symmetric around the 
glass intersection temperature. We used $ \alpha= 0.0055 {\rm Gauss} /K $ 
in the 
VG2-VG1 phase and  $ \alpha= 0.009 {\rm Gauss}/K $ in the BG2-BG1 phase.   
The left-hand y-axis denotes the scale for  $ B_{\rm dis}- \alpha T $,
the right-hand y-axis for  $ \partial B_{\rm dis}/\partial T  - 
\alpha  $. 
 } 
 \vspace*{0cm}
\end{figure}    

The upper picture in Fig.~5 shows the disorder part of the magnetic 
induction field $ B_{\rm dis} $ in the liquid phase 
for $  B=550 \; {\rm Gauss} $. In the lower picture we show  
the disorder part of the magnetic induction field $ B_{\rm dis} $ 
in the solid phase for 
$  B=180 \; {\rm Gauss} $. In correspondence to Beidenkopf {\it et al.} 
\cite{Beidenkopf1} we subtract to $ B_{\rm dis} $ 
a term linear in the temperature $ T $ to get a symmetrical 
curve around the glass transition temperature. 
We obtain a negligible jump 
$  \Delta \partial B_{\rm dis}(B,T) / \partial T $ over the glass 
transition line in the solid BG2-BG1 phase. This is in contrast to the 
jump  $  \Delta \partial B_{\rm dis}(B,T) / \partial T $ in the 
liquid high-temperature VG2-VG1 phase. By comparing the absolute values 
of this jump with the corresponding experimentally determined jump values 
$ \Delta \partial B(H,T) / \partial T $ determined in 
Ref.~\onlinecite{Beidenkopf1} our values are 
about one order of magnitude smaller. Note that for our theory 
$  \Delta \partial B_{\rm dis}(B,T) / \partial T $ is about 
$ 10^{-1} $ smaller in the BG2-BG1 phase than in the VG2-VG1 phase. 
This could not resolved within our numerics in Fig.~5. That this is true can  
be seen from the following scaling consideration
\begin{align} 
&  \Delta \frac{\partial}{\partial T} B_{\rm dis}(B,T)= -  
\frac{4 \pi}{v} 
 \Delta\Bigg[   \nonumber \\
& \int  ds 
\frac{\partial}{\partial B} 
\frac{\delta}{\delta \Delta(s)} \Delta f_{\rm var} 
\frac{\partial}{\partial T}\Delta(s) \! + \! \frac{\partial}{\partial T} 
\frac{\delta}{\delta \Delta(s)} \Delta f_{\rm var} 
\frac{\partial}{\partial B}\Delta(s) 
 \nonumber \\
&\!- \!
 \int \! ds ds' \! \left(\frac{\delta}{\delta \Delta(s)}
\frac{\delta}{\delta \Delta(s')}  \Delta f_{\rm var} \right) 
\frac{\partial}{\partial B} \Delta(s) \frac{\partial}{\partial T}  
\Delta(s')  \Bigg]
\nonumber \\
& 
 \sim  - \frac{4 \pi }{T B v} \left( \Delta f^{\rm kin}_{\rm var} + \Delta f^{\rm pot}_{\rm var} \; (Z^{(0)}_l \tilde{\Delta}(s_c))^2 \right) 
   \label{1270}
\end{align} 
where we used (\ref{1030}) in order to substitute terms containing 
$  \Delta f^{\rm pot}_{\rm var} $ to terms containing  
$ \Delta f^{\rm kin}_{\rm var} $. One can then see from analytic 
approximations but also numerical considerations that 
both terms in (\ref{1270}) are of almost equal value in the 
liquid phase VG2-VG1 but that the first kinetic term 
of the disorder free energy in (\ref{1270}) 
is much larger in the solid BG2-BG1 phase than the potential second part. 
Our numerics gives that the 
kinetic part of the disorder free energy $  \Delta f_{\rm var} $ 
in the BG2-BG1 phase is one order of magnitude smaller than 
in the VG2-VG1 phase.  

One source of the difference between the 
jump values of our theory and
the experimental numbers could beside the approximations we used 
in our theory also  the additional in-plane {\it ac} equilibrizing 
magnetic shaking field 
in the experiment of 
Beidenkopf {\it et al.} \cite{Beidenkopf3}. This shaking field is of 
the same magnitude as the {\it dc } magnetic field in z-direction. 
It is immediately clear from the results in Ref.~\cite{Beidenkopf1} 
as well as the theoretically and experimentally determined 
 results for an additional in-plane {\it dc} field instead of the 
$ {\it ac} $ field \cite{Schmidt1,Ooi1,Koshelev2} that the shaking 
field has only a small effect on the position of the first-order 
line and also the jump values $ \Delta S_l $ and $ \Delta B $. 
This can be understand by using the anisotropic scaling theory \cite{Blatter1} 
leading to an attenuation of the in-plane field by a factor 
$ \lambda_{ab}/ \lambda_{c} $. In contrast to this we obtain from Fig.~5 
that due to the smallness of the magnetic field $ B_{\rm dis} $ the 
shaking field can still have an effect on the jump value 
$\Delta  \partial B/ \partial T $ 
across the glass transition line BG2-BG1, VG2-VG1. Note that 
an in-plane  magnetic {\it dc} field can even put additional dislocations 
in the vortex lattice \cite{Koshelev2,Bolle1}.

\section{Summary}
In this paper, we have derived the phase diagram for
superconductors which have their phase transition lines  
at magnetic fields much smaller than $ H_{c2} $, i.e. $ B/H_{c2} < 
0.25 $ such as BSSCO. 
The model
 consists of the elastic degrees  of freedom
of the vortices with additional defect fields describing 
the defect degrees of freedom of the vortex lattice
in the most simple way.
For the  impurity potential we have restricted
ourselves 
 to weak pinning  $\delta T_c $ and $ \delta l $-correlated impurities
\cite{Blatter1}. 
This model was formerly used by us for describing the phase diagram 
of superconductors with a melting line near $ H_{c2} $ \cite{Dietel3}.  
The layered structure  of the superconductor, i.e. the Josephson coupling 
form between the layers, is not explicitly considered. We 
take this  special coupling only into account via the elastic moduli of the 
lattice and an experimentally  and analytically based 
decoupling scenario \cite{Glazman1, Daemen1, Goldin1}. 
In order to treat the 
impurity potential approximately 
we use a theory developed  first by M{\'e}zard and Parisi \cite{Mezard1} for 
random-manifolds. This is based on 
a variational approach to the free energy 
via a quadratic trial Hamiltonian. After stating our model in 
Section II we have discussed  the  M{\'e}zard-Parisi theory  
  of the vortex lattice system in Section III. 
The minimum   requirement for the trial free energy of the 
quadratic Hamiltonian leads to the saddle point equation (\ref{120}) 
where the stable solutions are full replica symmetric for $ {\cal D}(0) A < 
\sqrt{3}/2  $ 
in the fluid phase with  $ {\cal D}(0) A $ 
is defined in (\ref{122}) and (\ref{500}). Everywhere  else, the solutions 
are continuous replica symmetry broken.  
We expand these solutions to low temperatures $ {\cal D}(0) A > 
({\cal D}(0) A)_{\rm max} $ with (\ref{1220}). That the saddle point 
equation (\ref{120}) 
has no solution in general for $ {\cal D}(0) A 
\gtrsim ({\cal D}(0) A)_{\rm max} $ is shown in Appendix A.     
The M{\'e}zard-Parisi theory  is the lowest-order approximation of a 
more general perturbation theory known as variational 
perturbation theory. In Appendix B we show how to go beyond the 
lowest-order approximation for the vortex lattice system 
up to second-order where also in this case 
a saddle point solution is not existent.
 Motivated by good results of the variational 
perturbation treatment for the anharmonic oscillator we generalize in 
Section VI  the minimum requirement of the variational free energy 
determing the trial Hamiltonian to a more generalized 
principle of minimal sensitivity given in (\ref{1030}).
This leads to a second-order phase transition line located at the points 
in the $ H-T $ plane where the saddle point solutions cease  
to exist. 

As was discussed by us 
at the end of Section VI, for YBCO in contrast to BSCCO  
the magnitude of the lattice fluctuations 
near the melting line is smaller  than the correlation length of the 
impurity potential, i.e. 
$ (c_L a_0)^2 \ll \xi'^2 $ \cite{Dietel3}. This is the reason that 
one does not find a similar non-existence of  
saddle point solutions to the variational free energy, in certain 
regions of the $ H-T $ plane for YBCO as we obtain for BSCCO. 
This leads to the absence of the second-order phase 
transition line in YBCO.  
Further we note that Giamarchi and Doussal \cite{Giamarchi1}, 
who calculated the physics of the vortex lattice with pinning but 
without defects valid in the solid phase of real systems,  
did not 
find in their work the ceasing of saddle point solutions to the 
variational free energy in certain regions in 
the solid phase. The reason lies in the fact that they only consider 
small trial dimensionless gap functions 
$ \tilde{\Delta}(s) $ in their calculation. Also they did not use temperature 
softened elastic constants relevant for BSCCO \cite{Dietel2}.

The procedure described above leads to the following physical 
consequences for BSCCO. 
Due to the form of the elastic moduli in the deep $ H_{c2} $ 
region, we obtain two glass phase 
transitions of the depinning form. The first transition line of third-order 
is located in the fluid phase at high temperatures not far from $ T_c $.  
It is given by (\ref{1210}) identified as the depinning temperature of 
a coherently pinned vortex substring. It separates the full replica symmetric 
solution to the variational energy at high temperatures (VL phase) 
and a full replica symmetry broken solution at lower temperatures (VG1 phase). 
This transition corresponds to 
the glass transition in YBCO.  The transition line is located in the 
vicinity of the experimentally found $ T_x $ line \cite{Fuchs1}.  
The second transition is of  second-order (\ref{1220}) 
dividing the Bragg-glass and the vortex-glass phases in four regions. 
It separates a full replica symmetry broken saddle point solution 
of the variational free energy (VG1, BG1 phases) and a 
full replica symmetry broken turning point solution (VG2, BG2 phases).   
This transition line 
is a temperature depinning transition where a substring 
which is almost equally displaced due to disorder forming 
a plateau decouples 
from the impurities due to temperature fluctuations. 
We find that the derivate jump  of the magnetic 
induction field with respect to the temperature 
over this  glass transition line in the Bragg-glass phase 
is negligible in comparison to the jump in the liquid phase. 
We compare this line with the experimentally 
found second-order glass transition line by 
Beidenkopf {\it et al.} \cite{Beidenkopf1} 
located in the vicinity of our line. 
The jumps of the temperature  derivate of the magnetic induction field 
in the vortex-glass phase of our theory  
is about one order of magnitude lower in comparison to the 
experimental values of Beidenkopf {\it et al.} \cite{Beidenkopf1}.  
They obtain a similar value for the jumps in the Bragg-glass phase 
over the 
glass transition line. 
In comparison to the glass transition line separating the 
VG2-VG1 phases they found a softening of 
the jump in the vicinity of the glass transition line  
in the BG2-BG1 phase consistent with our findings.

We calculated the first-order melting transition line and its disorder 
induced continuation dividing the Bragg-glass phase BG2 and the 
vortex-glass phase VG2 by using an intersection criterium 
for  the low and the high-temperature  
expansion of the free energy.  
The whole theoretical determined phase diagram 
and the experimentally ones determined  by Beidenkopf {\it et al.} 
\cite{Beidenkopf1}  
is  shown in Fig.~3. Finally we compared 
the entropy jumps per layer and vortex, and also the magnetic induction 
jump over the first-order  line with the experimental findings of 
Zeldov {\it al.} \cite{Zeldov1}. This is shown in Fig.~4.

Summarizing, we have calculated the phase diagram of 
a vortex lattice model stated in \cite{Dietel2, Dietel3} for BSCCO 
without taking explicitly into account  the layered structure  
of the material. 
Although we found certain quantitative differences in the position 
of the experimental determined phase transition 
lines, the overall phase diagrams looks rather similar. 
Discrepancies are maybe due to the approximative evaluation 
of the theory and the layered structure of BSCCO.   

We would like to thank H.~Beidenkopf,  E.~H.~Brandt, and A.~Sudb{\o}
for useful discussions. This work was supported by 
Deutsche Forschungsgemeinschaft under grant KL 256/42-2.      

\begin{appendix} 
\section{General Proof of the non-existent 
of finite-step saddle 
point solutions for low temperatures} 
In this section we show that there exist no finite-step saddle point 
solution for the variational 
free energy $ \Delta f_{\rm var} $ in the range $
 {\cal D}(0) A \gtrsim  Z^{(0)}_l $ within first-order variational perturbation 
theory .
This was shown in the continuous and additionally 
 in the one-step case for the liquid phase in Section V. In order to derive 
this we use results derived in Section C in Ref.~\onlinecite{Dietel3}. 
We obtain for an $ R $-step replica symmetry breaking solution      
\begin{align}
& \frac{\sum_{i=1}^R \frac{1}{m^2_i} \left[S(\tilde{\Delta}_{m_i})-
S(\tilde{\Delta}_{m_{i-1}})\right]+ Z}
{\left(\sum_{i=1}^R\frac{
\Delta_{m_i} \! - \! \Delta_{m_{i-1}}}{m_i}
\right)^2 }   \nonumber \\
& =
  \frac{{\cal D}\left(2 \frac{k_B T}{v} g[\Delta_{m_R}]\right)
 }
 {\left(
2 \frac{k_B T}{v} 
{\cal D}'\left(2 \frac{k_B T}{v} g[\Delta_{m_R}]\right)
 \right)^2 
}    \label{a10}
\end{align} 
where we used that $ \Delta_0=0 $, $ m_{R+1} \equiv 1 $
and 
\begin{equation} 
S(\tilde{\Delta})\equiv - \int_0^{\Delta(s)} \! \! \! \! \!
d \Delta  \Delta \frac{d}{d \Delta} g(\Delta) 
\label{a20}  
\end{equation}  
It is shown in Ref.~\onlinecite{Dietel3} that $ Z > 0 $.
We have 
\begin{align} 
& S(\tilde{\Delta}) \approx 
\frac{1}{2} \tilde{\Delta}^{1/2}  \nonumber \\
& +   
 \left[ \log\left(1 + Z^{(0)}_l  \frac{\tilde{\Delta}}{2} \right) - 
\frac{Z^{(0)}_l   \frac{\tilde{\Delta}}{2} }{1+ Z^{(0)}_l   
 \frac{\tilde{\Delta}}{2}}\right]                     \label{a30} 
\end{align} 
in the liquid case and 
\begin{align} 
& S(\tilde{\Delta}) \approx  \frac{\sqrt{3}}{2} \frac{1}{8 \pi} 
\left( \frac{c_{44}^{(1)} a^2}{ c_{66} a_3^2} \right)  
 \Bigg\{
\frac{2}{3} \tilde{\Delta}^{3/2}   \nonumber \\
& +  \frac{4 Z^{(1)}_s}{(Z^{(0)}_l)^2}
 \left[ \log\left(1 + Z^{(0)}_l  \frac{\tilde{\Delta}}{2} \right) - 
\frac{Z^{(0)}_l   \frac{\tilde{\Delta}}{2} }{1+ Z^{(0)}_l   
 \frac{\tilde{\Delta}}{2}}\right]  \Bigg\}                 \label{a40} 
\end{align}  
for the solid. Next we use the inequalities
$ (\sum^R_{i=1} a_i)^2 \le R \sum_{i=1}^R a_i^2 $ for real number 
$ a_i, \ldots, a_R $ and further that 
$ \tilde{\Delta}_i \le \tilde{\Delta}_{i+1} $, 
$ \tilde{\Delta}_i^2/ 
\tilde{\Delta}^2_R \le 
S_{\tilde{\Delta}_i}
/ S_{\tilde{\Delta}_{R}} $, $ (1/m_i-1/m_{i+1})^2 \le 1/m^2_i -1/m_{i+1}^2 $ 
 resulting in  
\begin{equation}
\frac{S(\tilde{\Delta}_{m_R})} 
{R \,  \Delta^2_{m_R}}  \le 
  \frac{{\cal D}\left(2 \frac{k_B T}{v} g[\Delta_{m_R}]\right)
 }
 {\left[
2 \frac{k_B T}{v} 
{\cal D}'\left(2 \frac{k_B T}{v} g[\Delta_{m_R}]\right)
 \right]^2 
}  \,.   \label{a50}
\end{equation}  
This inequality can be only fulfilled for $ \sqrt{3} D(0) A/2  \lesssim  
4 R Z^{(0)}_l $ which is  a generalization of the  one-step replica symmetry 
breaking case in the fluid phase discussed below Eq.~(\ref{720}).

\section{Second-order variational perturbation 
expansion} 
In this Section we calculate the second-order expansion 
terms within variational perturbation theory (\ref{1000}). 
The aim is  to show that also to this order there are no saddle 
points of $ F_{\rm var} $ (\ref{1000}) with  $ D(0) A \gtrsim Z_{l}^{(0)} $ 
corresponding to (\ref{707}) and 
(\ref{712}) in the first-order case. We restrict ourselves to 
solutions of the saddle point equation with 
full replica symmetry  which were relevant in the 
first-order case according to Section IV.    

In the following, we will calculate first the expectation value 
of the disorder part of the replica Hamiltonian \cite{Dietel3}
\begin{equation}
H_{\rm dis}= \frac{-1}{2 k_B T} \sum_{\alpha, \beta} 
\sum_{{\bf x}} \delta_{x_3, x_3'}
\Delta(x_i+u_i^{\alpha}({\bf x})-
x_i-u_i^{\beta}({\bf x}')) \,,   \label{a500}
\end{equation}
in which $ \alpha $, $ \beta $ run over the replica indices. 
We allow only for onsite interactions which were justified in 
Ref.~\onlinecite{Dietel3} for YBCO. For the present compound 
this approximation is even 
more appropriate for BSCCO since $ \xi_{ab} \sim \xi'  \ll a $.  
With this disorder part (\ref{a500}) 
we obtain for $ H-H_{\rm trial} $ in (\ref{1000}) 
\begin{equation}  
H-H_{\rm trial} = H_{\rm dis}- \frac{v}{2}\sum_{{\bf x}}
\sum_{\alpha, \beta}  u_T^{\alpha}({\bf x})
\sigma_{\alpha \beta}
u_T^{\beta}({\bf x}) \,.  \label{a505}
\end{equation}
We now classify terms of 
higher-order variational perturbation theory 
in two groups. When expanding  
$ \langle (H-H_{\rm trial})^l \rangle_{\rm trial} $ in (\ref{1000}) 
we obtain first terms of the pure disorder Hamilton form  
$ \langle (H_{\rm dis})^l \rangle_{\rm trial,c} $ which we denote 
by $ \langle (H-H_{\rm trial})^l \rangle_{\rm trial,c,1} $. 
Second, there are monomials which contain at least one self-energy 
matrix factor $ \sigma_{\alpha \beta} $ in it denoted by 
 $ \langle (H-H_{\rm trial})^l \rangle_{\rm trial,c,2} $. 
These terms can be most
easily  treated by the square root trick \cite{Kleinertpath1}.
We now calculate first terms of the pure disorder Hamilton form. 

\subsection{Pure disorder terms in Hamiltonian} 

Within second-order variational perturbation theory, we obtain 
\begin{align}
& \langle (H_{\rm dis})^2 \rangle_{\rm trial}= 
\frac{1}{4 (2 \pi)^4 (k_B T)^2 }
\sum_{{\bf x},{\bf x}'} \sum_{\alpha, \beta, \gamma, \delta}  
\int d^2 q d^2q' 
  \nonumber \\
&  
\times  \hat{\Delta}
({\bf q}) \hat{\Delta}({\bf q}') 
\langle e^{i {\bf q}\cdot ({\bf u}^{\alpha}({\bf x}) -{\bf u}^{\beta}({\bf x}))+
i {\bf q}'\cdot ({\bf u}^{\gamma}({\bf x}') -{\bf u}^{\delta}
({\bf x}'))  } \rangle    \nonumber \\
& \approx \frac{1}{4 (2 \pi)^4 (k_B T)^2 }
\sum_{{\bf x},{\bf x}'} \sum_{\alpha, \beta, \gamma, \delta}
\int d^2 q d^2q' \hat{\Delta}({\bf q}) \hat{\Delta}({\bf q}') \nonumber \\
& 
\times e^{-\frac{k_B T}{v} 
\frac{q^2}{4} \left(G_{\alpha \alpha}(0)+
G_{\beta \beta}(0)- G_{\alpha \beta}(0)-G_{\beta \alpha}(0)\right)} 
\nonumber \\
& \times e^{-\frac{k_B T}{v} \frac{{q'}^2}{4} 
 \left(G_{\gamma \gamma}(0)+
 G_{\delta \delta}(0)- G_{\gamma \delta}(0)-G_{\delta \gamma}(0)\right)}
\nonumber \\
& \times  e^{- \frac{k_B T}{v}  {\bf q} \cdot \left({\bf G}_{\alpha \gamma}
 ({\bf x}-{\bf x}')+{\bf G}_{\beta \delta}({\bf x}-{\bf x}')-
 {\bf G}_{\beta \gamma}({\bf x}-{\bf x}')
 -{\bf G}_{\alpha \delta }({\bf x}-{\bf x}')\right){\bf q}'}   \,.  
\label{a510} 
\end{align}  
As before, we restrict ourselves 
 to the transversal part of the $ 2 \times 2 $ Green function 
$ {\bf G}({\bf x}) $ defined by 
$1/(2 \pi)^3  \int d^2 q d q_3 \;  ( {\bf q}_T \otimes {\bf q}_T)
 \; G({\bf q}) 
  e^{i q_i  x_i +i q_3  x_3} $. 
The replica sum in (\ref{a510}) is of the form 
$ \sum_{\alpha \beta \gamma \delta} 
F[G_{\alpha  \beta}, G_{\gamma  \delta}, 
 G_{\alpha  \gamma}, G_{\beta  \delta}, G_{\alpha \delta}, 
G_{\beta \gamma}] $   
where $F[\cdot] $ is some functional of the various Green functions. 
Since  $ G_{\alpha \beta} $ is some matrix within the Parisi algebra 
the functional $ F $ has the ultrametric property 
\cite{Dotsenko1, Temesvari1}.   
Following Temesv\'{a}ri {\it et al.}
\cite{Temesvari1}, we denote the size of the
Parisi blocks with  $ p_r $, $ r=1\ldots R$, where $ R $ is the maximum level
of replica symmetry breaking. We fix $ p_0=n $ and
$ p_{R+1}=1$, the latter being the size
of diagonal elements and $ n $ is the number of replica fields.
 The matrix elements $ \sigma_{\alpha \beta} $, that
belong to the $r$th level of replica symmetry breaking
are all equal to some number 
$ \sigma_r $, $ r=0,\ldots,R $. The replica overlap function  is defined by
$ \alpha \cap \beta =r $ when $ \sigma_{\alpha \beta}= \sigma_r $.

The fact that the Green function $ G_{\alpha \beta} $ is in the Parisi algebra 
implies that the Green function definitely depends on the overlap 
$ \alpha \cap \beta $ which we denote in the following by 
$ G_{\alpha \cap \beta} $. Furthermore, the operation $ \alpha \cap \beta $ 
on the replica indices has the ultrametric property. 
This means 
that whenever we may choose three replicas $ \alpha$, $\beta$, $ \gamma $, 
either all three of their overlaps are the same, i.e. 
$ \alpha \cap \beta = \alpha \cap \gamma = \beta \cap \gamma $, or one 
e.g. $ \alpha \cap \beta $ is larger than the other two. In the latter 
case the two  are equal, i.e.,  
$ \alpha \cap \beta > \alpha \cap \gamma = \beta \cap \gamma $.      
This means that of the three Green functions $ G_{\alpha \beta} $, 
$ G_{\alpha \gamma} $, and $ G_{\beta \gamma} $ only two are different.
Similarly, of the six Green functions  in $ F $ only three are different.
The various possible Green function combinations can be most easily 
determined by mapping these six Green functions onto the edges of a 
tetrahedron where the Green functions on the adjacent edges of a face  
must fulfill the ultrametric property. 

In the following, we restrict us to the 
leading term $ {\bf x}= {\bf x}' $ in (\ref{a510}). 
By carrying out the $ {\bf q} $, $ {\bf q}' $ integral we obtain 
\begin{widetext} 
\begin{align} 
& \!\!\!\! \! \! \!\!\!\!\!\!\!\!\!\!\!\!\!\!\!\!\!\!\!\!\! \!\!\!\!\!\!\!\!\! \langle (H_{\rm dis})^2 \rangle_{\rm trial,c} \approx  
(k_B T)^2 \frac{N }{4 }\sum_{\alpha, \beta, \gamma, \delta}    
{\cal D}^2(0)    \nonumber \\
& \times 
 \bigg( \frac{\xi'^4}{
 \left(\frac{k_B T}{v} 
  \left(g_{\alpha \alpha}-g_{\alpha \beta}\right)+\xi'^2 \right)
  \left(\frac{k_B T}{v} 
 \left(g_{\gamma \gamma }-g_{\gamma \delta}\right)+\xi'^2 \right)
 - \frac{1}{4}
 \left(\frac{k_B T}{v}\right)^2 \left(g_{\alpha \gamma} 
 + g_{\beta \delta}-g_{\beta \gamma }-g_{ \alpha \delta }\right)^2
} 
  \nonumber \\
&-  \frac{\xi'^4}{
 \left(\frac{k_B T}{v} 
  \left(g_{\alpha \alpha}-g_{\alpha \beta}\right)+\xi'^2 \right)
 \left(\frac{k_B T}{v} 
 \left(g_{\gamma \gamma }-g_{\gamma \delta}\right)+\xi'^2 \right)}  
\bigg)   \,.
\label{a530} 
\end{align} 
\end{widetext} 
The last subtracted term in (\ref{a530}) is due to the connectedness of 
$  \langle (H_{\rm dis})^2 \rangle_{\rm trial,c} $.
From (\ref{a530}) we obtain 
\begin{equation}
\sum_{\beta}  \sigma_{\alpha  \beta} =0    \,,        \label{a535} 
\end{equation}
being the same equation as in  the first-order 
variational perturbation theory case \cite{Mezard1, Dietel3}. 
We now restrict (\ref{a530}) to the Parisi algebra 
by carrying out the program outlined above leading 
for $ n \to 0 $ to  
 \begin{widetext}   
\begin{align} 
&   
\langle (H_{\rm dis})^2 \rangle_{\rm trial,c}=   
  n (k_B T)^2 N  {\cal D}^2(0)  \bigg\{
 \nonumber \\
 & 
2   \int ds_1 ds_2 ds_3  
 \left(-\Theta_{0,1}(s_1)+\delta(s_1-\sim)\right)  
 \left(-\Theta_{s_1,1}(s_2)+\delta(s_2-\sim)-s_2\delta(s_2-s_1)\right) 
\left(-\Theta_{s_2,1}(s_2)+\delta(s_3-\sim)-s_3\delta(s_3-s_2)\right)   
 \nonumber \\
 & 
 \bigg(\frac{\xi'^4 }{
 \left(\frac{k_B T}{v} 
  \left( \tilde{g}-g_{s_1} \right)+\xi'^2 \right)
 \left(\frac{k_B T}{v} 
 \left(\tilde{g}-g_{s_2} \right)+\xi'^2 \right)
 - \frac{1}{4}
 \left(\frac{k_B T}{v}\right)^2 \left(g_{s_2}-g_{s_3}\right)^2
 }   -\frac{\xi'^4 }{
  \left(\frac{k_B T}{v} 
  \left( \tilde{g}-g_{s_1} \right)+\xi'^2 \right)
 \left(\frac{k_B T}{v} 
  \left(\tilde{g}-g_{s_2} \right)+\xi'^2 \right) 
  }   \bigg)                  \nonumber \\
&  +  \int ds_1 ds_2 ds_3  
 \left(-\Theta_{0,1}(s_1)+\delta(s_1-\sim)\right)  
 \left(-\Theta_{0,s_1}(s_2)+\delta(s_2-s_1)\right) 
 \left(-\Theta_{0,s_2}(s_3)+s_3\delta(s_3-s_2)\right) 
 \nonumber \\
&   
 \bigg( \frac{\xi'^4 }{
  \left(\frac{k_B T}{v} 
  \left( \tilde{g}-g_{s_3} \right)+\xi'^2\right)^2
 - \frac{1}{4}
 \left(\frac{k_B T}{v}\right)^2 \left(g_{s_2}-g_{s_1}\right)^2 } 
 - \frac{\xi'^4 }{
 \left(\frac{k_B T}{v} 
 \left( \tilde{g}-g_{s_3} \right)+\xi'^2\right)^2 }
\bigg)  \bigg\}  \,.
\label{a540} 
\end{align} 
\end{widetext} 
Here $ g_{s} $  
is the momentum integrated Green function of 
$ G_s $ according to Eq.~(\ref{70}). 
We define $ \delta(s_i-\sim) $ by the functional 
$ \int ds_i \delta(s_i-\sim ) H[g_{s_i}] =H(\tilde{g}) $ where $ H $ is 
some functional of the integrated Green function $ g_{s_i} $ and 
$ \tilde{g} \equiv  g_{\alpha \alpha} $.  

For calculating the saddle point equation up to second
order variational perturbation theory corresponding to (\ref{120}) 
the derivate 
 $  (\delta/ \delta g_{s}) \langle (H_{\rm dis})^2 
 \rangle_{\rm trial,c} $ 
is relevant which should be added with an appropriate 
factor to the right hand side of Eq.~(\ref{120}). In order to derive 
this equation we first give the variational free energy 
$ F_{\rm var}/N $ within second-order variational perturbation theory 
denoted by $ f_{\rm var,2} $   
 \begin{equation}
  f_{\rm var,2}= f_{\rm var,1}-
 \frac{1}{2 (k_B T)} \langle  (H_{\rm dis})^2 
 \rangle_{\rm trial,c}   \label{a550}  
 \end{equation} 
where  $ f_{\rm var,1} $ corresponds to the variational energy 
within first-order variational perturbation theory given 
in (\ref{54}) and (\ref{60}), i.e. 
$ f_{\rm var,1}= f_{\rm var}(0)+\Delta f_{\rm var,1} $   
where $ \Delta f_{\rm var,1} $ is a modification  of 
 $ \Delta f_{\rm var} $  specified in (\ref{54}) according to 
\begin{equation} 
\Delta f_{\rm var,1}= P_1 \Delta f^{\rm kin}_{\rm var}+ P_2  
\Delta f^{\rm pot}_{\rm var}
\,.
\label{a555} 
\end{equation} 
The additional   prefactors $ P_i $ are modifications due to 
second-order perturbational expansion from terms 
proportional to 
$ \langle (H -H_{\rm trial})^2  \rangle_{\rm trial,c} $ (\ref{a505}) 
containing at least one factor $ \sigma_{\alpha \beta} $. 
The constants $ P_i $ lying between $ 1/2 $ and $ 3/2 $ 
will be determined in the next subsection.   

Carrying out the variation of $ f_{\rm var,2} $ 
with respect to $ g_s $ we obtain 
 \begin{equation}
 P_3 \sigma(s) = -  2 \,  \frac{k_B T}{v} \;
 P_4 {\cal D}' \left(2 B[\Delta(s)] \right)-  
 \frac{\delta}{ \delta g_{s}} \frac{\langle (H_{\rm dis})^2 
\rangle_{\rm trial,c} }{(k_B T)^2}   \,.
 \label{a560}
 \end{equation}
 The calculation of  $ (\delta/ \delta g_{s}) \langle (H_{\rm dis})^2 
 \rangle_{\rm trial,c} $ is tedious but straight-forward. 
Due to lack of space, we do not state the result here. 

In order to discuss the sign of 
  $ \langle (H_{\rm dis})^2 
 \rangle_{\rm trial,c} $ and  
 $ (\delta/ \delta g_{s}) \langle (H_{\rm dis})^2 
 \rangle_{\rm trial,c} $ we repeat the form of the Green functions 
in the Parisi algebra \cite{Mezard1}
\begin{align}
&  \tilde{g}-g_s= \frac{1}{(2 \pi)^3} 
 \int d^2kdk_3 \bigg[\frac{1}{G_0^{-1}({\bf k},k_3)+\Delta(1)}    \nonumber \\
& +
 \int^1_s ds' 
\frac{\sigma'(s)}{\left(G_0^{-1}({\bf k},k_3)+\Delta(s)\right)^2}
\bigg]
                       \label{a570}
\end{align} 
and 
\begin{align} 
&  \tilde{g}= \frac{1}{(2 \pi)^3} 
 \int d^2kdk_3 \; G_0({\bf k},k_3)        \nonumber \\
& \times  \left[1
 +\int_0^1 ds \frac{1}{s^2} 
\frac{\Delta(s)}{G_0^{-1}({\bf k},k_3)+\Delta(s)}\right]      \,.
\label{a580}  
\end{align}  
 By using (\ref{a550}), (\ref{a560}) we obtain    
  \begin{align}
 &  P_3 \sigma'(s) =  - \sigma'(s)  
    \left(\frac{2 k_B T}{v}\right) g'(\Delta(s))  \label{a590} \\
& \times \bigg( \frac{2 k_B T }{v} 
 P_4 {\cal D}'' \left(2 B[\Delta(s)]\right) 
+ D_s 
 \frac{\delta}{\delta g_{s}}\frac{ 
\langle (H_{\rm dis})^2  \rangle_{\rm trial,c}}{(k_B T)^2} \bigg) 
\nonumber 
 \end{align}
 corresponding to (\ref{525}) in the first-order case.
Here, we have used the differential operator    
\begin{equation}
D_s  
= \left( \sigma'(s)  
    \left(\frac{2 k_B T}{v}\right) g'[\Delta(s)]\right)^{-1}
 \frac{\partial}{\partial s }  \,.  \label{a600}  
\end{equation}  
Dividing (\ref{a590}) by $ \sigma'(s) $ and forming the 
derivate  with respect to $ s $, we obtain   
\begin{align} 
&   \left(\frac{2 k_B T}{v}\right)^2
  \frac{g'[\Delta(s)]^3}{g''[\Delta(s)]} \; 
\bigg( \frac{2 k_B T}{v} P_4 {\cal D}'''\left(2 B[\Delta(s)] \right)
\nonumber \\
& \qquad \qquad +D_s^2 
 \frac{\delta}{\delta g_{s}} 
\frac{\langle (H_{\rm dis})^2  \rangle_{\rm trial,c}}{(k_B T)^2} 
\bigg)  = s      \,.   \label{a610} 
\end{align} 
In contrast to the first-order results  (\ref{525}), (\ref{528}) the 
second-order variational perturbation equations (\ref{a590}), (\ref{a610}) 
give no longer local algebraic equations for $ B[\Delta(s)] $ and 
$ \Delta(s) $ but integral equations involving  both quantities  for 
different $s $.

From (\ref{a540}) we obtain that  
$ \delta/\delta g_s \langle (H_{\rm dis})^2  \rangle_{\rm trial,c}$  
depends through $ g_s $ on $ s $. 
One can show after a  tedious but straight-forward analysis 
that 
\begin{align} 
& \langle (H_{\rm dis})^2 
  \rangle_{\rm trial,c}  >   0     \,,  \label{a620}   \\
& \frac{\delta}{ \delta g_{s_c}} 
\langle (H_{\rm dis})^2   \rangle_{\rm trial,c}   >  0 
\,, \label{a621}    \\
& 
 D_{s_s} \frac{\delta}{ \delta g_{s_c}} 
\langle (H_{\rm dis})^2   \rangle_{\rm trial,c}  >   0 \,. \label{a622}
\end{align} 
Here $ s_c $ is defined such that $ \sigma'(s)= 0 $ for $ s_c \le s \le 1 $.
As in the first-order case, $ s_c $ can be determined by (\ref{a590}) 
for $ s=s_c $ with $ B[\Delta(s_c)]= g[\Delta(s_c)] $. Then we obtain 
by the help of (\ref{a622}) that equation (\ref{a590}) is not 
solvable at $ s=s_c $ for small temperatures. More precisely 
we find that (\ref{a590}) is not solvable for 
$ {\cal D}(0)A \gtrsim Z_{l}^{(0)} $ 
by using     
$  ({\cal D}''(2 B[\Delta(s_c)])/v)^{-1} 
   D_{s_c}(\delta/\delta g_{s_c}) 
\langle (H_{\rm dis})^2 \rangle_{\rm trial,c} /(k_B T)^3 \sim 
{\cal D}(0)  A/  Z_{l}^{(0)} $. 

\subsection{Terms containing at least one factor  
$ \sigma_{\alpha \beta}$} 
Next, we consider contributions to second-order variational perturbation 
 expansion $ \langle (H-H_{\rm trial})^2 \rangle_{\rm trial,c} $ (\ref{1000}) 
 containing at least one factor 
$ \sigma_{\alpha \beta} $. As described in the textbook 
Ref.~\onlinecite{Kleinertpath1} for  the case of the anharmonic oscillator,  
these terms can be best derived with  the help of the square root trick. 
In our system this trick consists in substituting 
$ \Delta(s) $ in $ \Delta f_{\rm var} $ of Eq.~(\ref{54}) by 
$ (1-k)\Delta(s) $ denoted by $ \Delta f_{\rm var}(k) $. The 
$ \langle (H-H_{\rm trial})^2 \rangle_{\rm trial,c,2}   $ terms 
for  $ k=0 $  containing at least one factor 
$ \sigma_{\alpha \beta} $ are then given by 
\begin{align} 
& \langle (H-H_{\rm trial})^2 \rangle_{\rm trial,c,2} \nonumber \\
& = -2 
 k_B T \left( \frac{1}{2}\frac{\partial}{\partial k^2}   
\Delta f^{\rm kin} _{\rm var}(0)+ \frac{\partial}{\partial k}   
\Delta f^{\rm pot} _{\rm var}(0)\right) \,.   \label{a700}       
\end{align}      
This leads 
to the contributions in  $ \Delta f_{\rm var,2} $ (\ref{a550}) and the 
saddle point equation (\ref{560}) 
which are a factor $ (\Delta \partial/\partial \Delta)
 g(\Delta) /g(\Delta) $ or 
$ (\Delta \partial/\partial \Delta)^{1+m}
 g(\Delta) / (\Delta \partial/\partial \Delta) 
g(\Delta) $   where $ m=1,2 $ smaller than the leading 
contributions. By using (\ref{120}), (\ref{130}) we obtain only 
non-negligible contributions  to 
$ f_{\rm var,2} $ (\ref{a550}) or the 
saddle point equation (\ref{a560}),  i.e. $ P_i \not=1 $, for 
$ \tilde{\Delta} \ll 1/(Z^{(0)}_l)^{4/3} $ in the 
fluid phase and $ \tilde{\Delta} \ll 1/(Z^{(1)}_s)^2 $ in the solid 
phase. We point out that $ \tilde{\Delta}_{\rm max} $ (\ref{708}) in the fluid 
phase and (\ref{713}) in the solid phase in first-order 
variational perturbation theory is much larger than these  
$ \tilde{\Delta} $ values. Note that we obtain also corrections 
$ P_i \not=1 $ in the regime $ \tilde{\Delta} \gg 1/Z^{(0)}_l $ 
much larger than $ \tilde{\Delta}_{\rm max} $. 

Thus, we consider the regime $ \tilde{\Delta} \ll 1/(Z^{(0)}_l)^{4/3} $ in the 
fluid phase, and $ \tilde{\Delta} \ll 1/(Z^{(1)}_s)^2 $ in the solid 
phase. Here, we obtain prefactors $ P_i $ in (\ref{a550}), (\ref{a560}) 
which differ in general for both phases. 
We obtain 
\begin{eqnarray}
& P^{T \to 0}_{1} = \frac{11}{8} \;, \;  
 P^{T \to \infty}_{1}= \frac{7}{8}  \; , \;      
 P^{T \to 0}_{2}= 1  \;, \; 
P^{T \to \infty}_{2}= \frac{1}{2} \,,  \nonumber  \\     
&  P^{T \to 0}_{3}= \frac{11}{8} \, , \,       
      P^{T \to \infty}_{3}= \frac{7}{8} \,,\, 
 P^{T \to 0}_{4}= P^{T \to \infty}_{4}= \frac{1}{2} \,. \qquad  \label{a710}         \end{eqnarray} 
\end{appendix}


\begin{thebibliography}{99}
\bibitem{Bem2}
One can show that the quadratic disorder 
fluctuations in the  
single pinning regime \cite{Blatter1} 
are much smaller than $ a^2 $ in the interesting regime near the melting line, 
i.e. $ u^2(0,a_3) \ll a^2 $ with $ a_3 $ given 
by (\ref{22}).
\bibitem{Bem1} 
This can be shown by using the methods of Section II B in Ref.
\onlinecite{Blatter1} using a dynamical approach to the 
pinning of a single vortex.
\end{thebibliography}
\end{document}